\documentclass[iop, twocolumn]{emulateapj}
\usepackage{url}
\usepackage{graphicx}
\usepackage{natbib}
\usepackage{textcomp}
\usepackage{epsfig}
\usepackage{amsmath}
\usepackage{booktabs}
\usepackage{hyperref}

\slugcomment{Draft version \today}

\shortauthors{Awasthi A. K., Sylwester B., Sylwester J., Jain R.}
\shorttitle{Energy-dependent Thermal characteristics of the DEM[T] distribution}

\begin{document}
\title{Thermal characteristics and the differential emission measure distribution during a B8.3 flare on July 04, 2009}

\author{Arun Kumar Awasthi\altaffilmark{1,*}}
\affil{Astronomical Institute, University of Wroclaw, Wroclaw, Poland}
\email{*arun.awasthi.87@gmail.com/awasthi@astro.uni.wroc.pl}

\author{Barbara Sylwester\altaffilmark{2} and Janusz Sylwester\altaffilmark{2}}
\affil{Solar Physics Division, Space Research Centre, Polish Academy of Sciences,
Wroclaw, Poland}

\and

\author{Rajmal Jain\altaffilmark{3}}
\affil{Kadi Sarva Vishwavidyalaya, Gandhinagar, Gujarat, India}

\begin{abstract}
We investigate the evolution of differential emission measure distribution (DEM[T]) in various phases of a B8.3 flare, which occurred on July 04, 2009. We analyze the soft X-ray (SXR) emission in 1.6-8.0 keV range, recorded collectively by Solar Photometer in X-rays (SphinX; Polish) and Solar X-ray Spectrometer (SOXS; Indian) instruments. We make a comparative investigation of the best-fit DEM[T] distributions derived by employing various inversion schemes viz. single gaussian, power-law, functions and Withbroe-Sylwester (W-S) maximum likelihood algorithm. In addition, SXR spectrum in three different energy bands viz. 1.6-5.0 keV (low), 5.0-8.0 keV (high) and 1.6-8.0 keV (combined) is analyzed to determine the dependence of the best-fit DEM[T] distribution on the selection of energy interval. The evolution of DEM[T] distribution, derived using W-S algorithm, reveals the plasma of multi-thermal nature during the rise to the maximum phase of the flare, while of isothermal nature in the post-maximum phase of the flare. Thermal energy content is estimated considering the flare plasma to be of 1) iso-thermal and 2) multi-thermal nature. We find that the energy content during the flare, estimated from the multi-thermal approach, is in good agreement with that derived using the iso-thermal assumption except during the maximum of the flare. Further, (multi-) thermal energy estimated employing low-energy band of the SXR spectrum result in higher values than that derived from the combined-energy band. On the contrary, the analysis of high-energy band of SXR spectrum lead to lower thermal energy than that estimated from the combined-energy band.
\end{abstract}


\maketitle

\section{Introduction}
Solar flare is one of the most energetic phenomena occurring in the atmosphere of our Sun, releasing typically $\mathrm{10^{27}}$-$\mathrm{10^{32}}$ ergs of energy in $\sim$ $10^3$ s. This immense energy release is understood to be powered by the magnetic energy via the process of magnetic reconnection \citep{Shibata1999, Jain2011a, Choudhary2013, Aschwanden2014, Dalmasse2015}. A typical M-class solar flare can be observed across almost entire electromagnetic spectrum \citep{Benz2008, Fletcher2011}. Therefore, various energy release processes occurring at various heights in the solar atmosphere can be probed by the investigation of the observed multi-wavelength flare emission.

X-ray emission during solar flares mainly originates from the corona and upper chromosphere. Moreover, X-ray emission recorded during a flare can serve as the best probe of studying various plasma processes of thermal and non-thermal character \citep{Li2005, Saint-Hilaire2005, Jain2008, Awasthi2014}. Low-energy X-ray emission ($<$ 10 keV), also known as soft X-ray (SXR), is understood to be originated in the process of free-free, free-bound and bound-bound emission due to collision of charged particles (mostly electrons) having thermal (Maxwell-Boltzmann) distribution. On the other hand, high energy X-ray emission (hard X-rays) is known to be produced as a consequence of thick-target bremsstrahlung of non-thermal electron beam with the dense plasma in the chromosphere \citep{Brown1971, Kulinova2011}. Moreover, SXR emission during a flare is understood to be produced by multi-thermal plasma \citep{Aschwanden2007, Jain2011b, Sylwester2014, Aschwanden2015}.

The study of thermal characteristics of the flare plasma is made by the inversion of observed X-ray spectrum through postulating an empirical functional form of differential emission measure distribution (DEM[T]). Although DEM[T] plays a key role in deriving thermal characteristics and in turn energetics of the flare plasma, it is less accurately known owing to the fact that inversion of observed radiation needs to be performed which is very ill-posed problem \citep{Craig1976}. Moreover, several DEM[T] schemes which postulate certain functional dependence of DEM on T viz. single gaussian, bi-gaussian, power-law etc.  have been proposed (see \citet{Aschwanden2015} for exhaustive list of schemes). Further, a Withbroe-Sylwester (W-S) maximum likelihood DEM inversion algorithm has been established by \citet{Sylwester1980} where the functional form of DEM[T] is not a-priori defined. In this regards, a comparative survey of the aforesaid DEM schemes in the form of derived thermal characteristics of the flare plasma is very necessary provided application of various inversion schemes result in similar outcome. In addition, an inevitable restriction on deriving the complete thermal characteristics of the flare plasma is posed by the availability of observations from different instruments in certain specific energy-bands only. Thermal emission can be best studied by measuring the X-ray spectrum in typically 1-12 keV energy band. Observations in aforesaid energy band with high spectral and temporal cadence is very difficult to achieve from a single instrument due to huge difference of flux during a flare across the said energy band.

Therefore, we examine temperature dependence of differential emission measure from the analysis of multi-instrument data for a B8.3 flare which occurred on July 04, 2009. As the flare selected for the analysis is the only event common between Solar Photometer in X-Rays (SphinX; a Polish instrument) and Solar X-ray spectrometer (SOXS; an Indian instrument), combined data-set provides a unique opportunity for exhaustive study of the complete thermal characteristics of a small flare.  Both the instruments make use of Si PIN detectors for observing solar atmosphere in X-ray waveband. Section \ref{sec:obs} presents observations used for the present study and the specification of respective instruments. In Section \ref{sec:data-analysis}, we present the study of DEM[T] distribution derived by employing different inversion schemes and it's dependence on the selection of energy band of input SXR spectrum. In Section \ref{sec:e-th}, thermal energetics of the flare, estimated from the parameters derived from various schemes, is presented. Section \ref{sec:sum-conc} is comprised of the summary and conclusions.

\section{Observations}
\label{sec:obs}
\begin{figure*}[!htbp]
 \begin{center}
   \includegraphics[height=0.9\textwidth, angle=270]{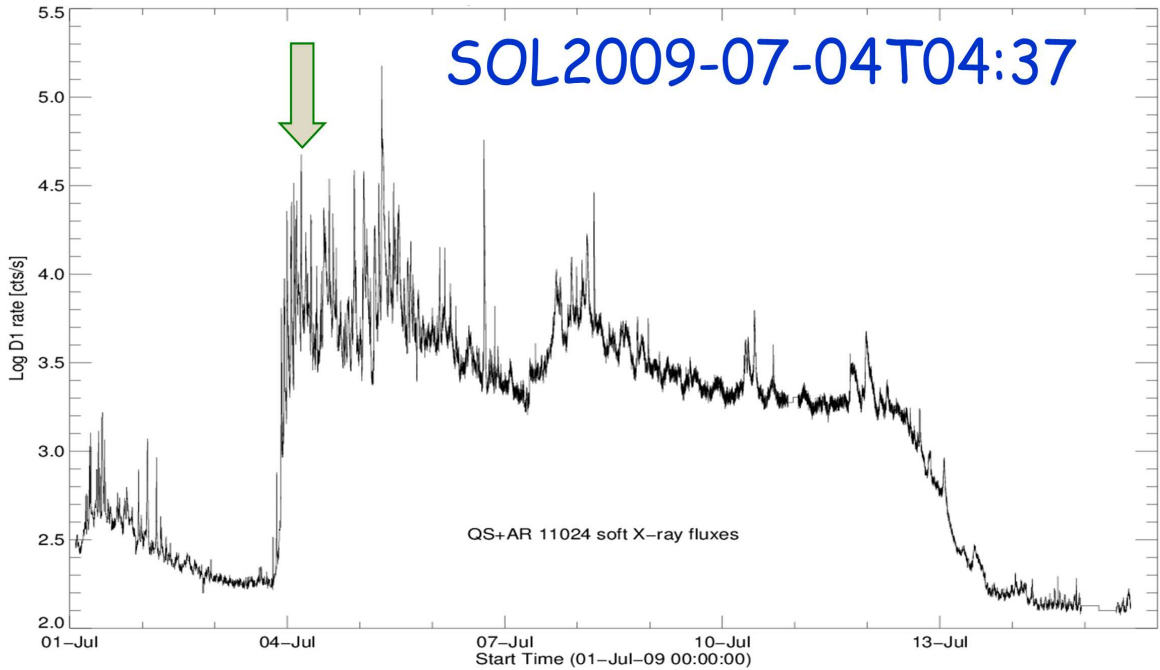}
   \caption{X-ray light-curve of the solar corona, dominated by the emission from a single active region AR 11024 with the flaring emission atop as seen by SphinX. Flare SOL2009-07-04T04:37, selected for the present study, is shown by an arrow.}
   \label{light-curve-all-sphinx}
\end{center}
\end{figure*}

We investigate a B8.3 intensity class flare which occurred on July 04, 2009 in an active region AR11024. AR 11024 appeared on the disk on July 3, 2009 and rotated off the disk on July 15, 2009. More than 500 flares or small brightenings have been observed with SphinX mission on the soft X-ray light curve during that time. Figure \ref{light-curve-all-sphinx} shows the temporal evolution of X-ray emission recorded by SphinX during the aforesaid period.

The SOL2009-07-04T04:37 flare, selected for the present study, is the only event observed in common by SphinX and SOXS missions due to the fact that SOXS mission usually observed the Sun in X-rays for only 2-3 hours in a day. We analyze the X-ray spectra in 1.6-5.0 keV and 5.0-8.0 keV energy bands, recorded from SphinX and SOXS missions, respectively. We briefly discuss the data and respective instruments' specifications as following:

\subsection{Solar Photometer in X-Rays (SphinX) mission}
We analyze X-ray spectra in 1.6-5.0 keV (hereafter low-energy band) from the SphinX instrument \citep{Gburek2011, Sylwester2012, Gburek2013}. SphinX, a spectrophotometer designed to observe solar corona in soft X-rays, was flown on-board the Russian \textit{CORONAS-PHOTON} satellite on January 30, 2009. SphinX employed three Si PIN diode detectors to record X-rays in the energy range $\sim$ 1.2-15.0 keV. The temporal and spectral cadence of SphinX observations are as good as 6 $\mu$s and 0.4 keV, respectively. Detailed information regarding the observations, procedure for calibration and data warehouse may be referred from \citet{Gburek2013} and from the SphinX instrument homepage\footnote{\url{http://156.17.94.1/sphinx_l1_catalogue/SphinX_cat_main.html}}.

\subsection{Solar X-ray Spectrometer (SOXS) mission}
X-ray spectra in 5.0-8.0 keV (hereafter high-energy band) during the flare are obtained by the Solar X-ray Spectrometer (SOXS) instrument \citep{Jain2005, Jain2008}. SOXS employed two semiconductor devices, viz. silicon (Si) PIN detector for recording X-ray observations in the energy range 4-25 keV while Cadmium Zinc Telluride (CZT) detector for that in the energy range 4-56 keV. The energy resolution of the Si detector is $\sim$ 0.8 keV while that for CZT detector is $\sim$ 1.7 keV. The temporal cadence of the observations obtained from both the detectors is 3 sec during the quiet and gradual phase of the flare. However, on-board automated algorithm allowed to record the observations with 100 msec cadence during the rise to peak phase of the flare. The data obtained during the entire observing-span (May 2003- April 2011) of SOXS mission and the analysis procedures are available on the instrument homepage\footnote{\url{https://www.prl.res.in/~soxs-data/}}. In the present study, we employ the observations obtained from Si detector in view of it's better energy resolution and sensitivity in comparison to that of CZT detector.

Left panel of the Figure \ref{x-ray-sphinx-soxs-goes-lc} presents the evolution of X-ray emission as observed by SphinX (top row) and SOXS (bottom row) missions during the flare in various energy bands plotted with different colors. Intensity curves shown by black and red colors represent the X-ray emission recorded by SphinX in 1.6-3.0 keV and 3.0-5.0 keV, respectively. Further, X-ray emission in 5.0-7.0 keV and 7.0-8.0 keV, drawn by blue and green colors, respectively are obtained from the SOXS.

It may be noted from the Figure \ref{x-ray-sphinx-soxs-goes-lc} that SOXS receives higher background than that seen by SphinX. On the contrary, a comparison of the count rates recorded by SphinX and SOXS in 4-6 keV, the energy band commonly covered by both the instruments, revealed that SOXS observations are lower by a factor of $\sim$2.5. This may be attributed to the systematic difference of sensitivities between the two instruments. \citet{Mrozek2012} reported higher flux in SphinX 3-8 keV energy band in comparison to the observations obtained in the same energy range from \textit{RHESSI} mission by a factor varying in the range of 2-6. On the other hand, the comparison of SOXS and \textit{RHESSI} observations in 6-12 keV energy band has been performed by \citet{Caspi2010}, which resulted in the agreement of the spectra, obtained from both the instruments, within 5\%-10\%. In this study, we prepared combined-data by applying the aforesaid `empirical normalization factor' in the records obtained from SOXS. On the other hand, we consider the flux recorded by SphinX as the true flux owing to the fact that this is the only instrument available to observe X-ray emission in the energy less than 4 keV during the flare. Therefore, the difference of inter-instrument sensitivity and hence normalization factor in the aforesaid energy range can't be established. However, in order to study the effect of this approximation, we have also carried out the investigation of emission measure distribution by applying the inverse normalization factor on the SphinX records while retaining the SOXS counts as such. We discuss the effect of both the aforesaid cases on the thermal energy estimates as presented in the Section \ref{sec:e-th}.

\subsection{Geostationary Operational Environmental Satellite (\textit{GOES})}
\begin{figure*}[!htbp]
 \begin{center}
   \includegraphics[height=0.9\textwidth, angle=90]{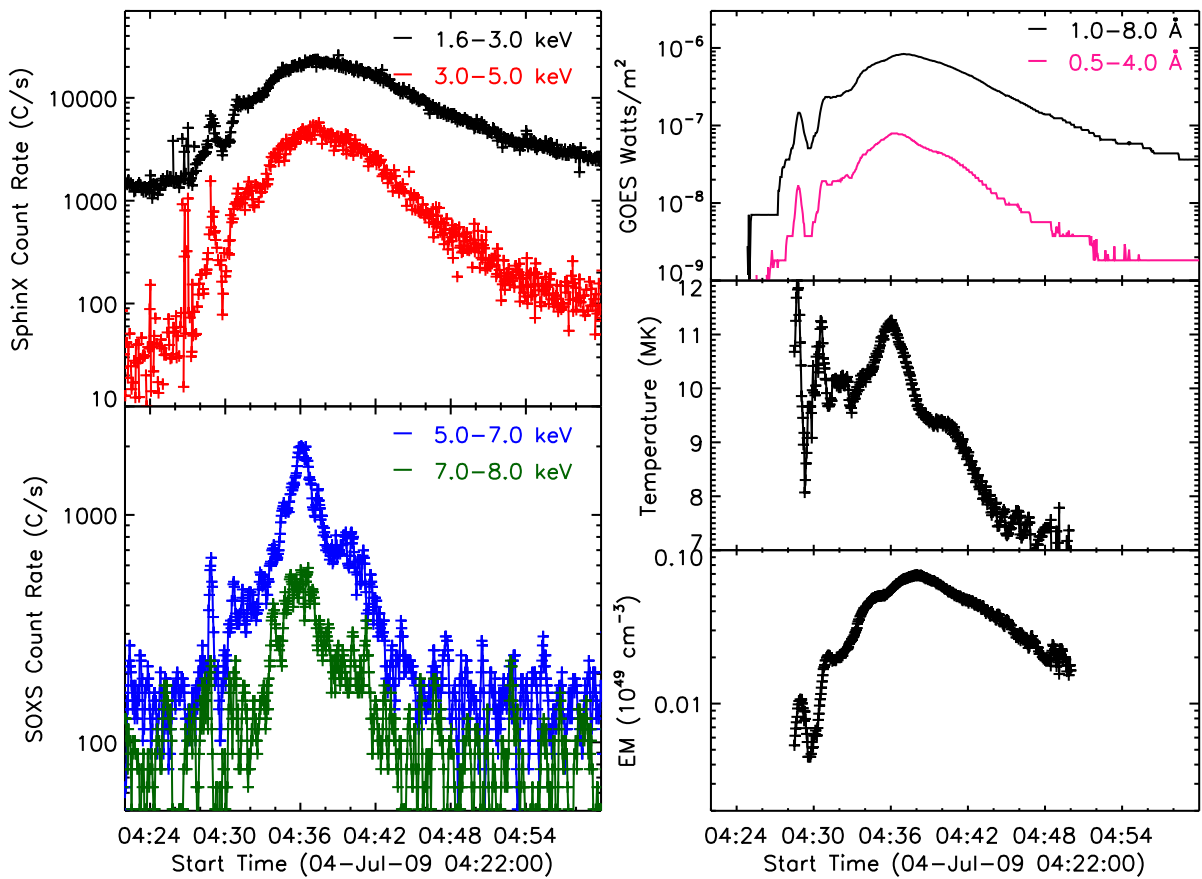}
   \caption{Left panel: Temporal evolution of X-ray count rate in 1.6-3.0 keV and 3.0-5.0 keV obtained from SphinX (top row) and that in 5.0-7.0 keV and 7.0-8.0 keV, as recorded by SOXS (bottom row). Right panel: \textit{GOES} flux in 1.0-8.0 $\mathrm{\AA}$ and 0.5-4.0 $\mathrm{\AA}$ (top row), temperature (middle row) and emission measure (bottom row).}
\label{x-ray-sphinx-soxs-goes-lc}
\end{center}
\end{figure*}

Geostationary Operational Environmental Satellite (\textit{GOES}) refers to a series of satellites dedicated to observe X-ray emission from Sun-as-a-star in two wavelength bands viz. 1.0-8.0 $\mathrm{\AA}$ and 0.5-4.0 $\mathrm{\AA}$. Right column of the Figure \ref{x-ray-sphinx-soxs-goes-lc} shows the background subtracted flux in 1.0-8.0 $\mathrm{\AA}$ and 0.5-4.0 $\mathrm{\AA}$ bands plotted by black and red colors, respectively. We treat the observations averaged during 04:20-04:25 UT as the background. Temperature and emission measure (EM) estimated from the flux-ratio technique adopted for \textit{GOES} data are also plotted in the middle and bottom rows of the right panel of Figure \ref{x-ray-sphinx-soxs-goes-lc}, respectively. It may be noted that temperature is found to be varying in the range of 7-12 MK while the EM in the range of 0.003-0.08 $ \times  10^{49} \mathrm{cm^{-3}}$. We use these T \& EM estimates to calculate the thermal energy during the flare (cf. Section \ref{sec:e-th}).

\subsection{Morphology of the flare in Extreme Ultraviolet (EUV) emission}
\begin{figure*}[!htbp]
\begin{tabular}{cccc}
\epsfig{file=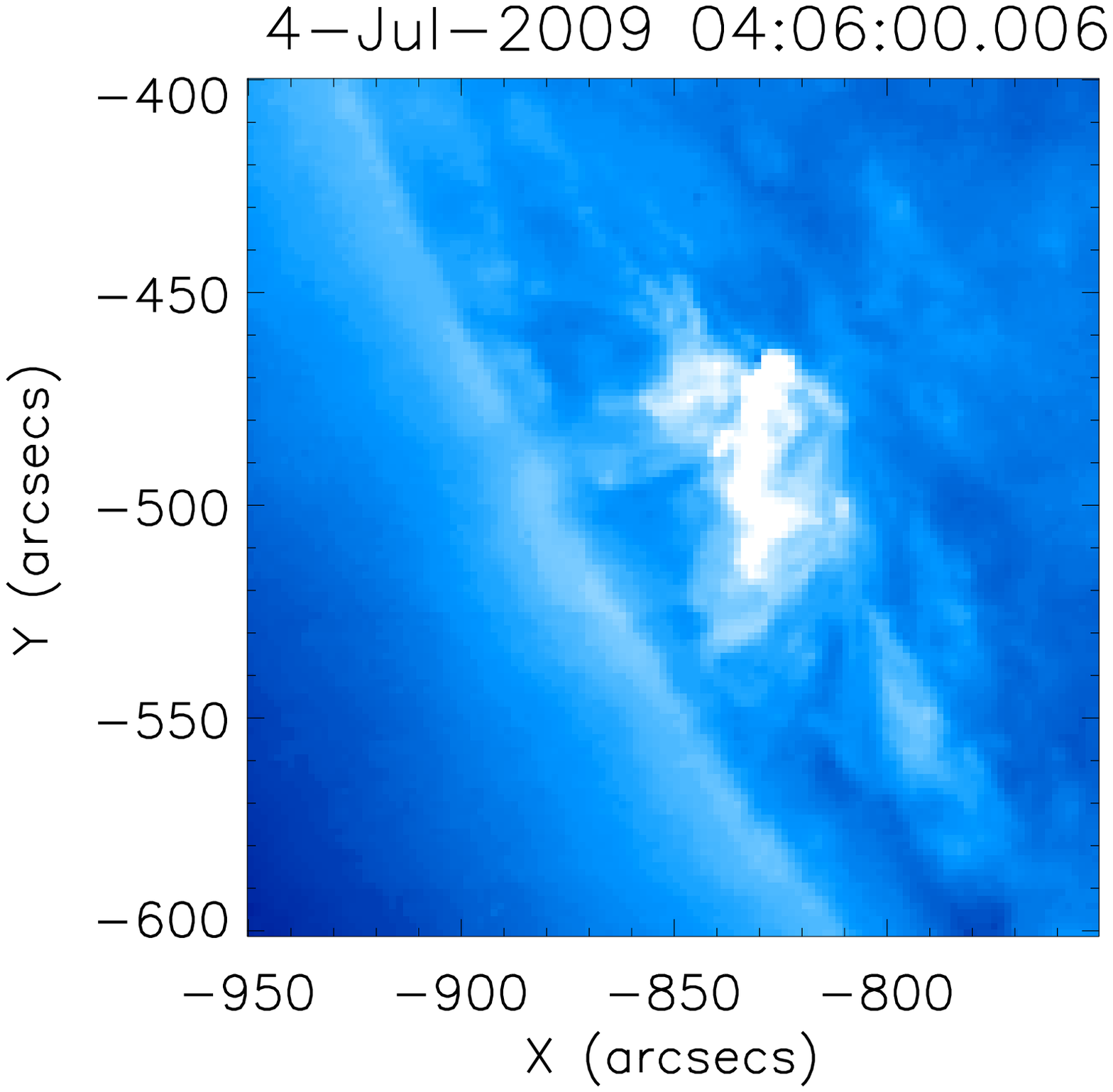, angle=90, width=0.25\textwidth, trim=20 150 20 150} &
\epsfig{file=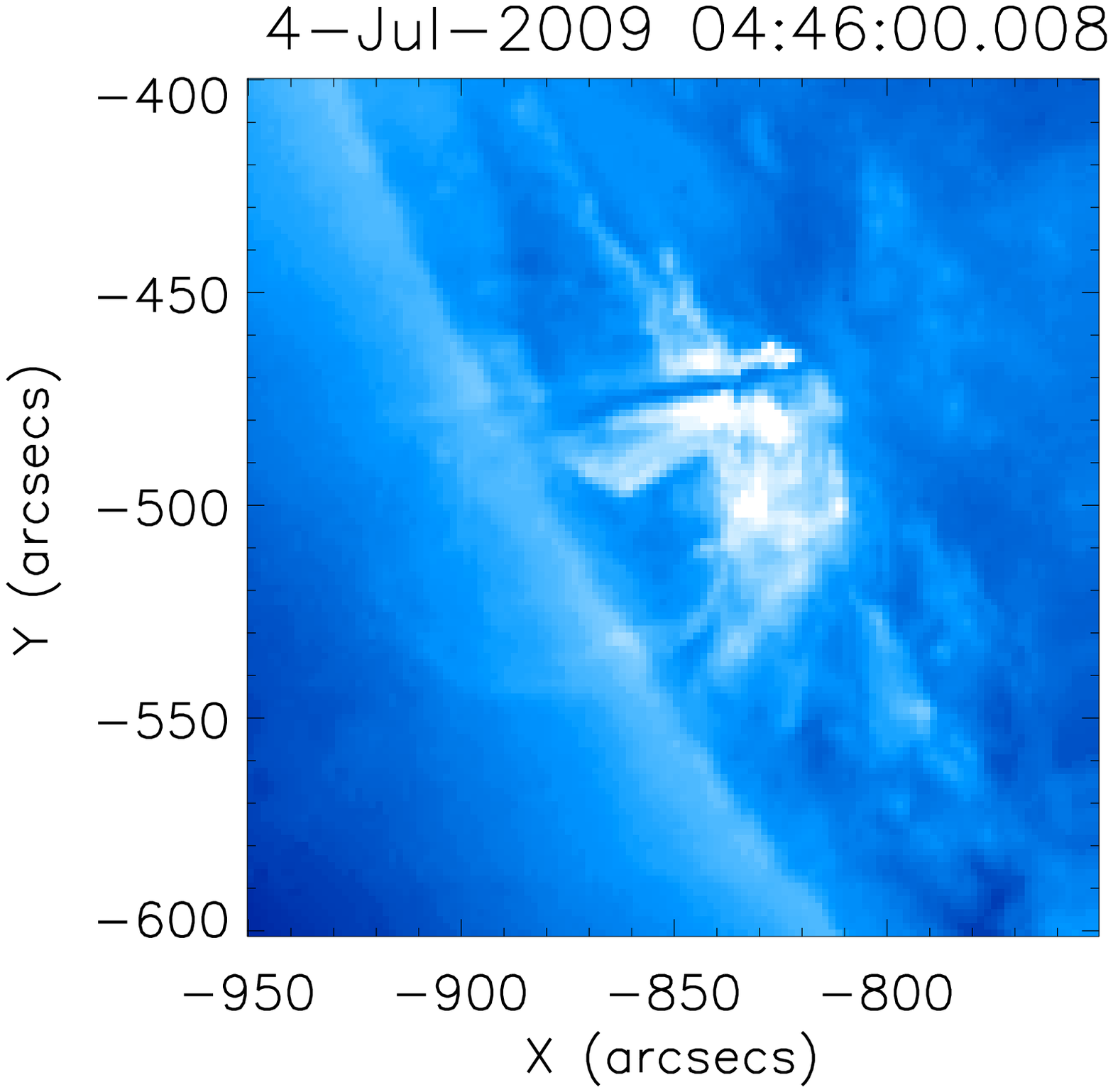, angle=90,  width=0.25\textwidth, trim=20 150 20 150} &
\epsfig{file=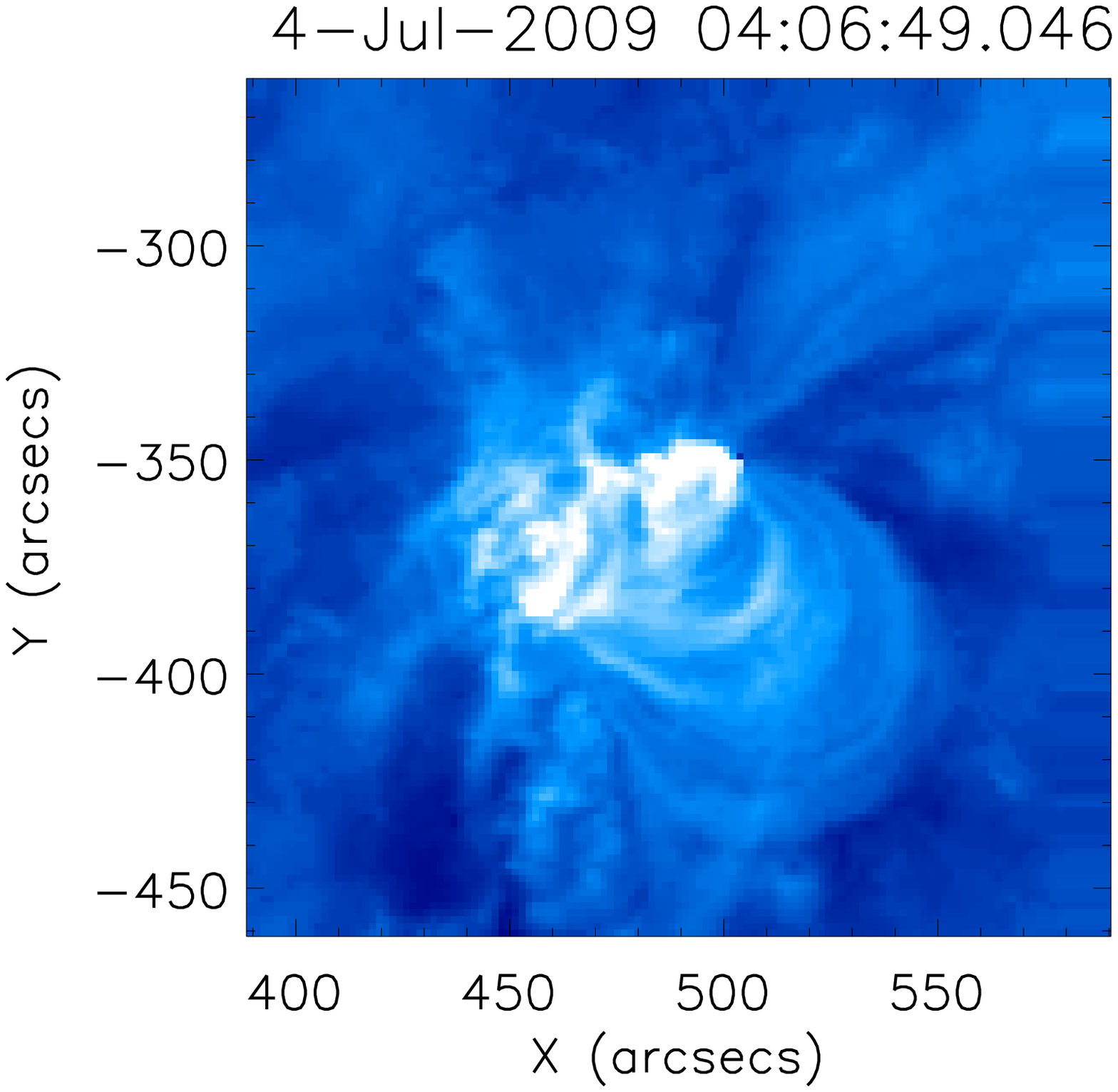, angle=90, width=0.25\textwidth, trim=20 150 20 150} &
\epsfig{file=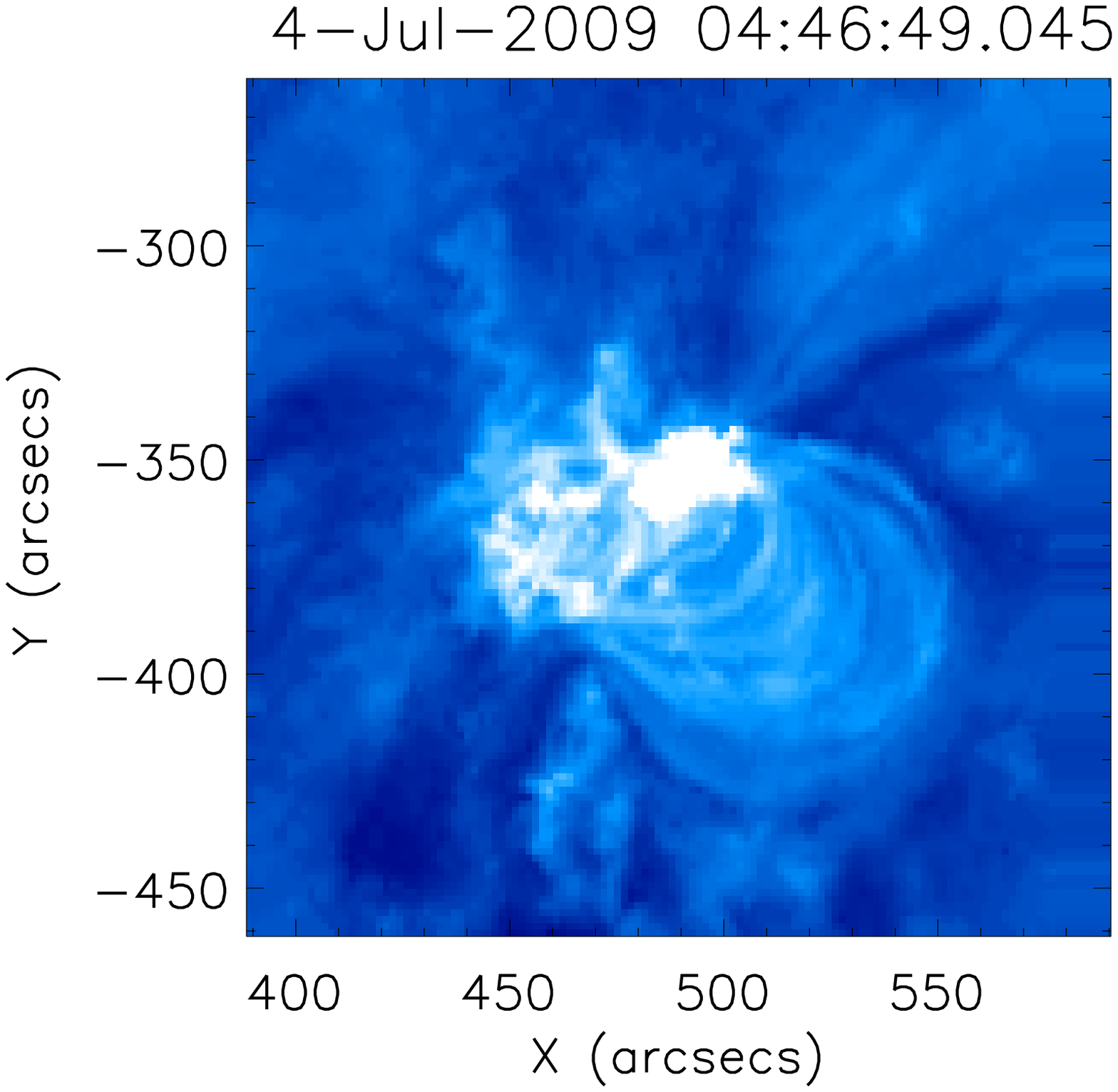, angle=90,  width=0.25\textwidth, trim=20 150 20 150} \\
\epsfig{file=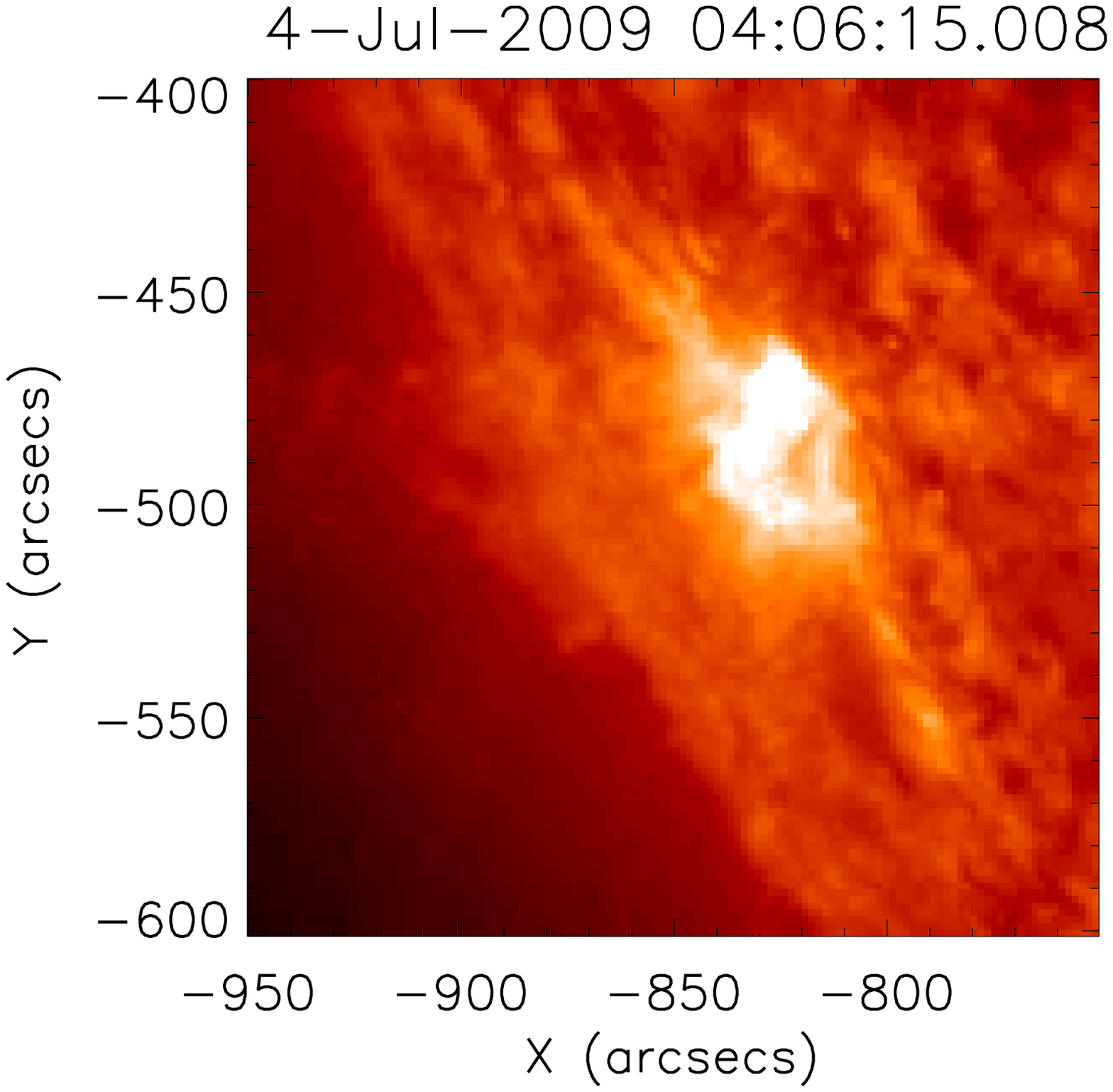, angle=90, width=0.25\textwidth, trim=20 150 20 150} &
\epsfig{file=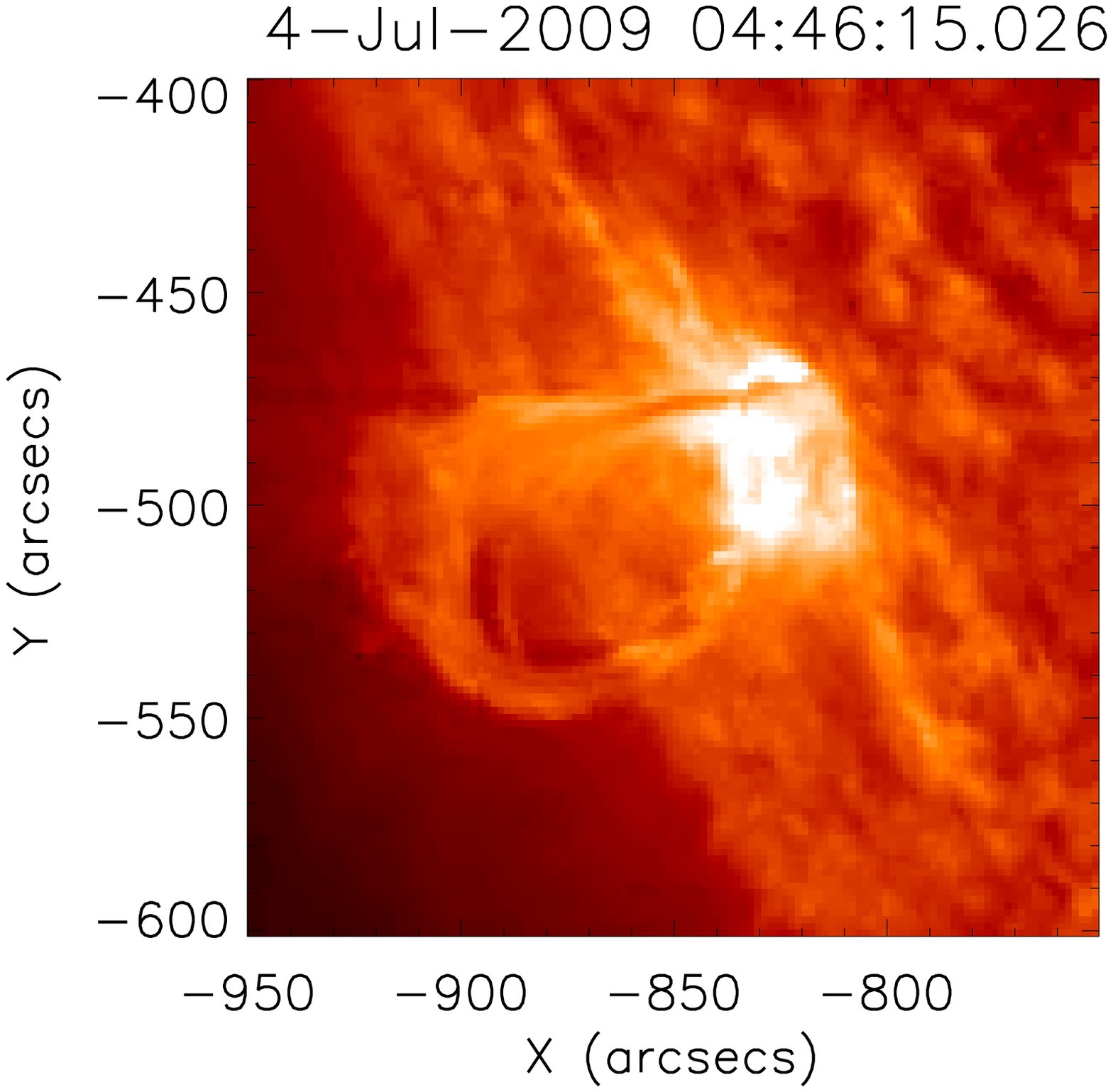, angle=90, width=0.25\textwidth, trim=20 150 20 150} &
\epsfig{file=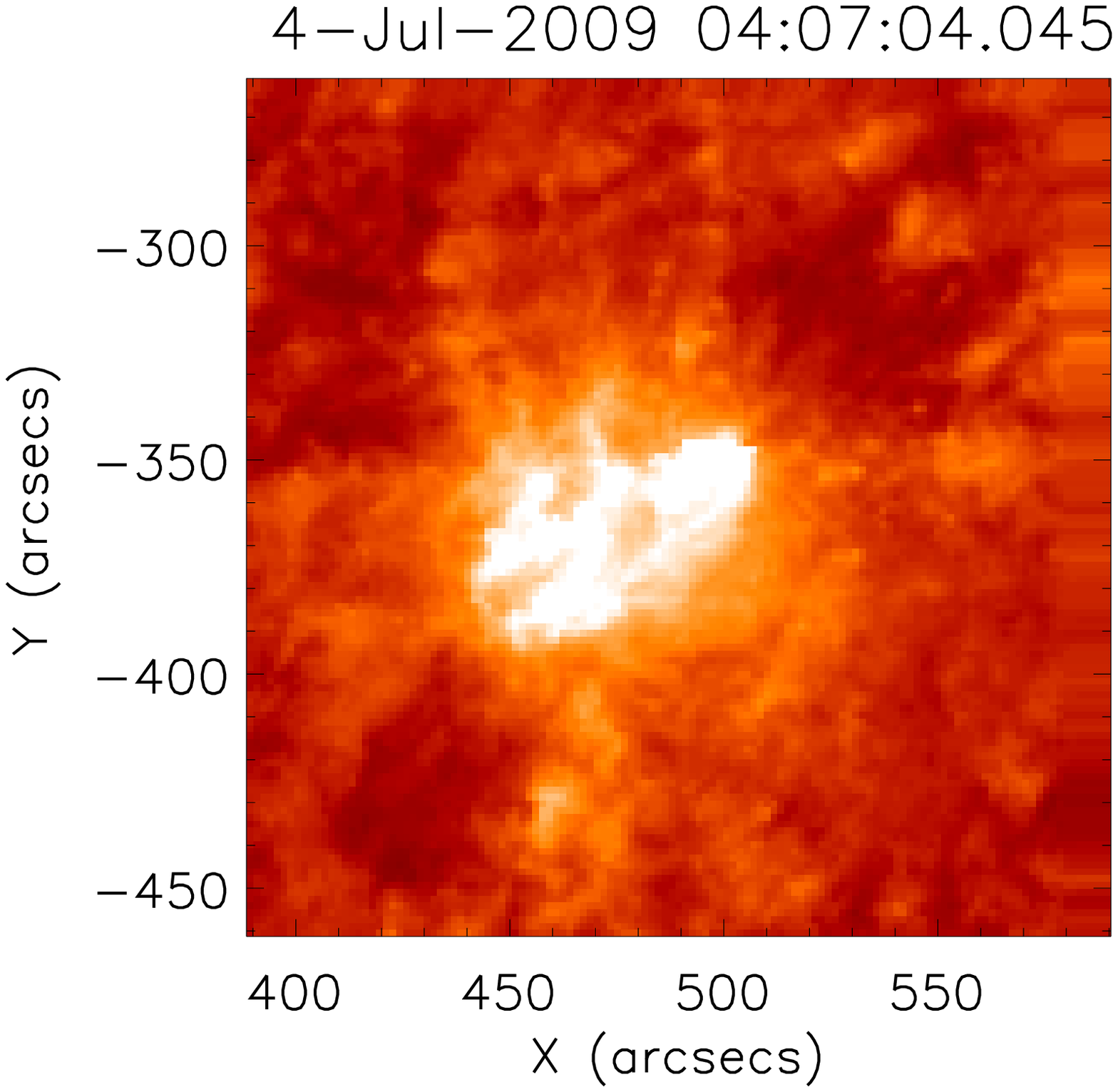, angle=90, width=0.25\textwidth, trim=20 150 20 150} &
\epsfig{file=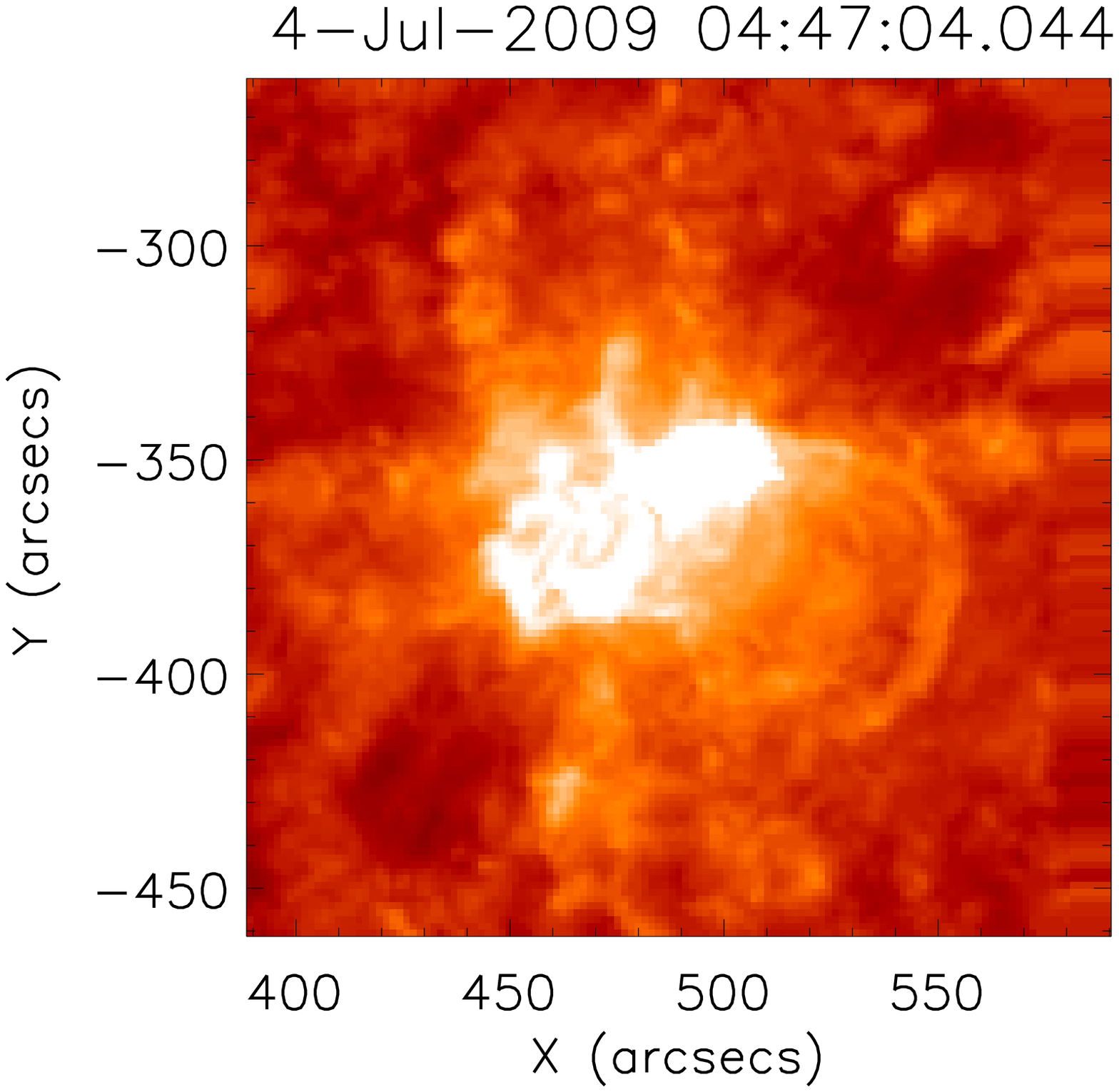, angle=90, width=0.25\textwidth, trim=20 150 20 150}
\end{tabular}
\caption{Time sequences of images in 171 $\mathrm{\AA}$ (top row), and 304 $\mathrm{\AA}$ (bottom row) obtained by \textit{STEREO} twin-satellite during the flare. Images in the first two columns correspond to the side view of the flare and obtained from \textit{STEREO-A} while that in the rest two columns present the line-of-sight view of the flare, as seen from \textit{STEREO-B}.}
\label{stereo-twin}
\end{figure*}

Temporal and morphological evolution of the flaring region is studied from the observations obtained from EUV Imaging Telescope (EIT) \citep{Delaboudiniere1995} on-board \textit{Solar and Heliospheric Observatory (SOHO)}. Moreover, images in 171, 284 and 304 $\mathrm{\AA}$ wavelength, recorded by \textit{STEREO} twin-satellites, are also processed. In Figure \ref{stereo-twin}, we present the morphological evolution of the flaring region in 171 and 304 $\mathrm{\AA}$ during the flare as obtained from \textit{STEREO-A} \& \textit{B}.

From the time sequence of the EUV images presented in Figure \ref{stereo-twin}, we note that although the flare event considered is of a small B8.3 intensity class, it is associated with an eruption. The study of the eruption is out of the scope of the aim of this paper. We estimate the volume of the emitting region from the EUV images to derive thermal energetics of the flare as presented in section \ref{sec:e-th}.

\section{DEM[T] distribution from the application of various inversion schemes}
\label{sec:data-analysis}
In order to study thermal characteristics of the flare plasma, we make an exhaustive investigation of evolution of DEM[T] relationship employing the X-ray spectra observed from SphinX and SOXS. We explore the dependence of DEM[T] distribution, which is derived by employing various inversion schemes on the SXR spectra of various energy bands viz. 1.6-5.0 keV (low-energy), 5.0-8.0 keV (high-energy) and 1.6-8.0 keV (hereafter combined-energy). This study aims to understand the dependence of the best-fit DEM[T] representing a selective part of the SXR emission, which in turn presents the consequence of restrictions posed by the co-temporal observations recorded in separate energy bands from different instruments viz. SphinX and SOXS. In this study, we employ the DEM inversion schemes which postulate (1) single-gaussian and, (2) power-law functional relationship of DEM with T. In addition, we also employ (3) a well-established Withbroe-Sylwester (W-S) maximum likelihood inversion algorithm which is independent of a-priori assumption of a functional form of DEM[T]. In the following, we present the thermal characteristics of X-ray emission during the flare as derived by applying aforesaid inversion schemes.

\subsection{DEM varying as a single-gaussian function over temperature}
We investigate the best-fit DEM[T] distributions, obtained by employing the scheme of single-gaussian functional dependence of DEM on $T$, on the observed SXR spectrum in the low, high and combined energy bands. However, firstly, we also employ this DEM scheme on synthesized model multi-thermal spectrum. Below we discuss the aforesaid two cases.

\subsubsection{DEM[T] distribution of a synthesized model multi-thermal spectrum}
\begin{figure}[!htbp]
   \includegraphics[height=0.45\textwidth, angle=90]{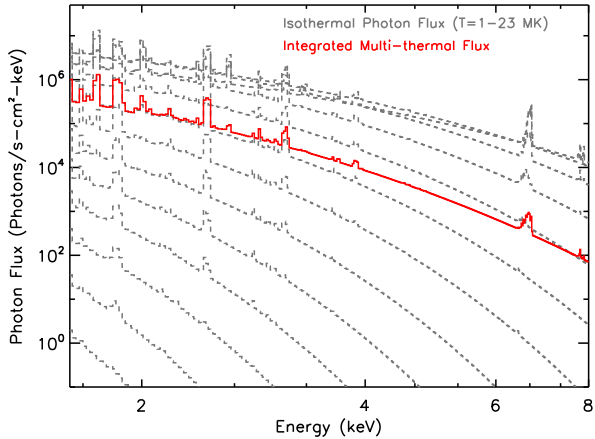}
   \caption{The spectra plotted in grey color (dotted) represent the iso-thermal photon flux corresponding to the temperature range 1-23 MK with the interval log T=0.1 MK. The plot drawn with red color represents the integrated (multi-thermal) photon flux.}
   \label{theory-multithermal-spectra}
\end{figure}

\begin{figure*}[!htbp]
\begin{tabular}{cc}
\epsfig{file=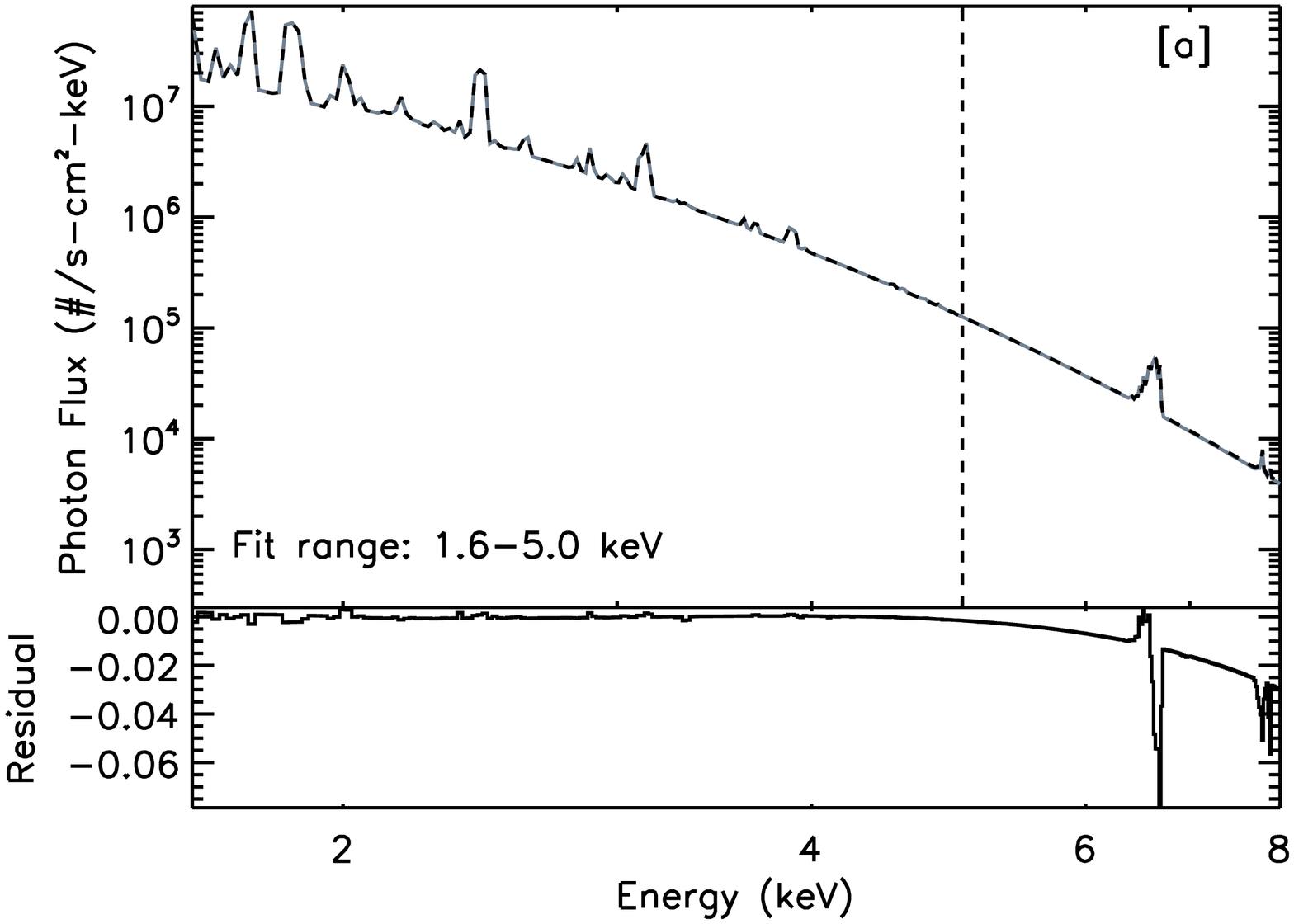, angle=90, height=0.31\textwidth, clip, trim=0 0 0 0} &
\epsfig{file=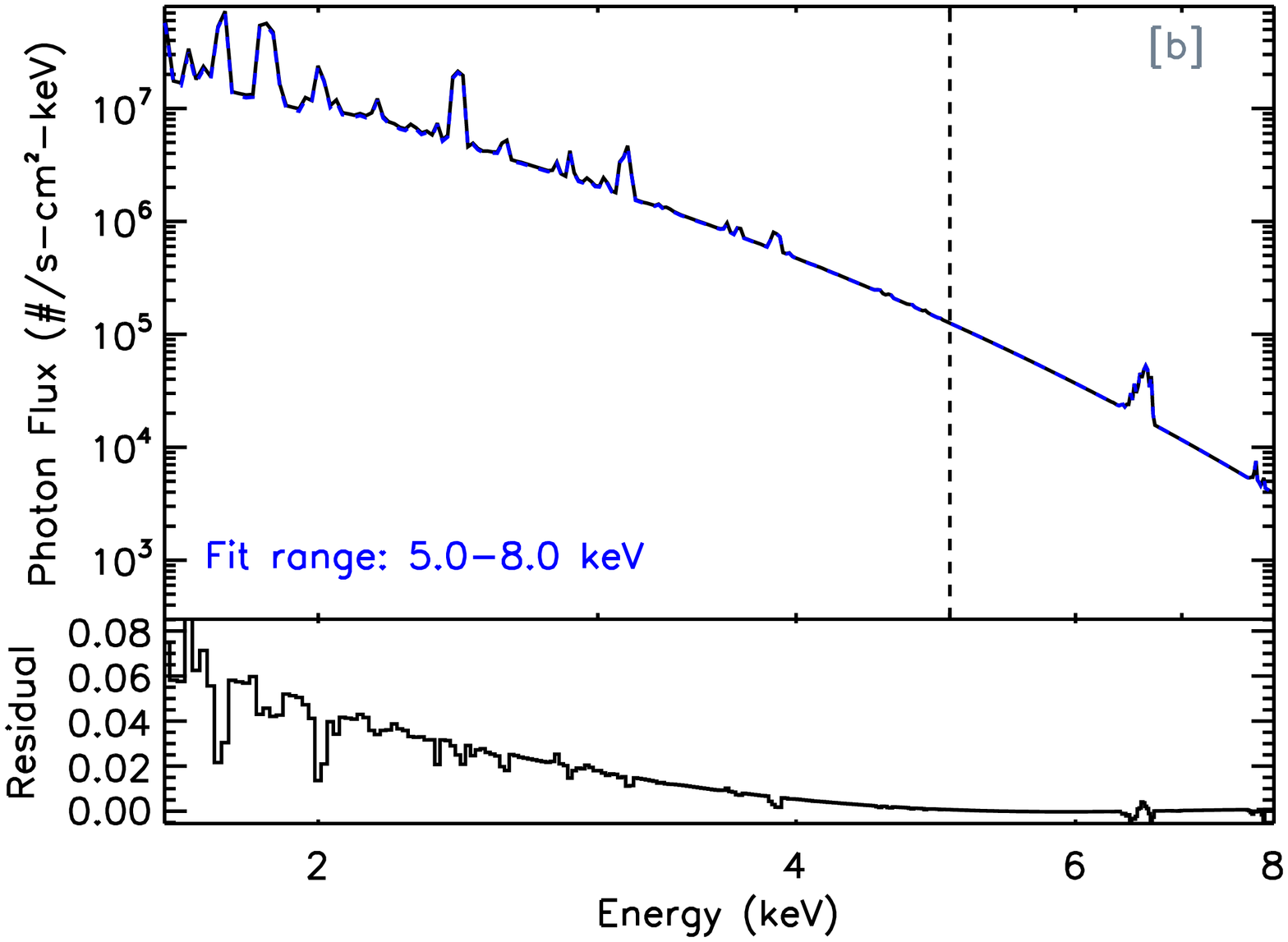, angle=90, height=0.31\textwidth, clip, trim= 0 0 0 0} \\
\epsfig{file=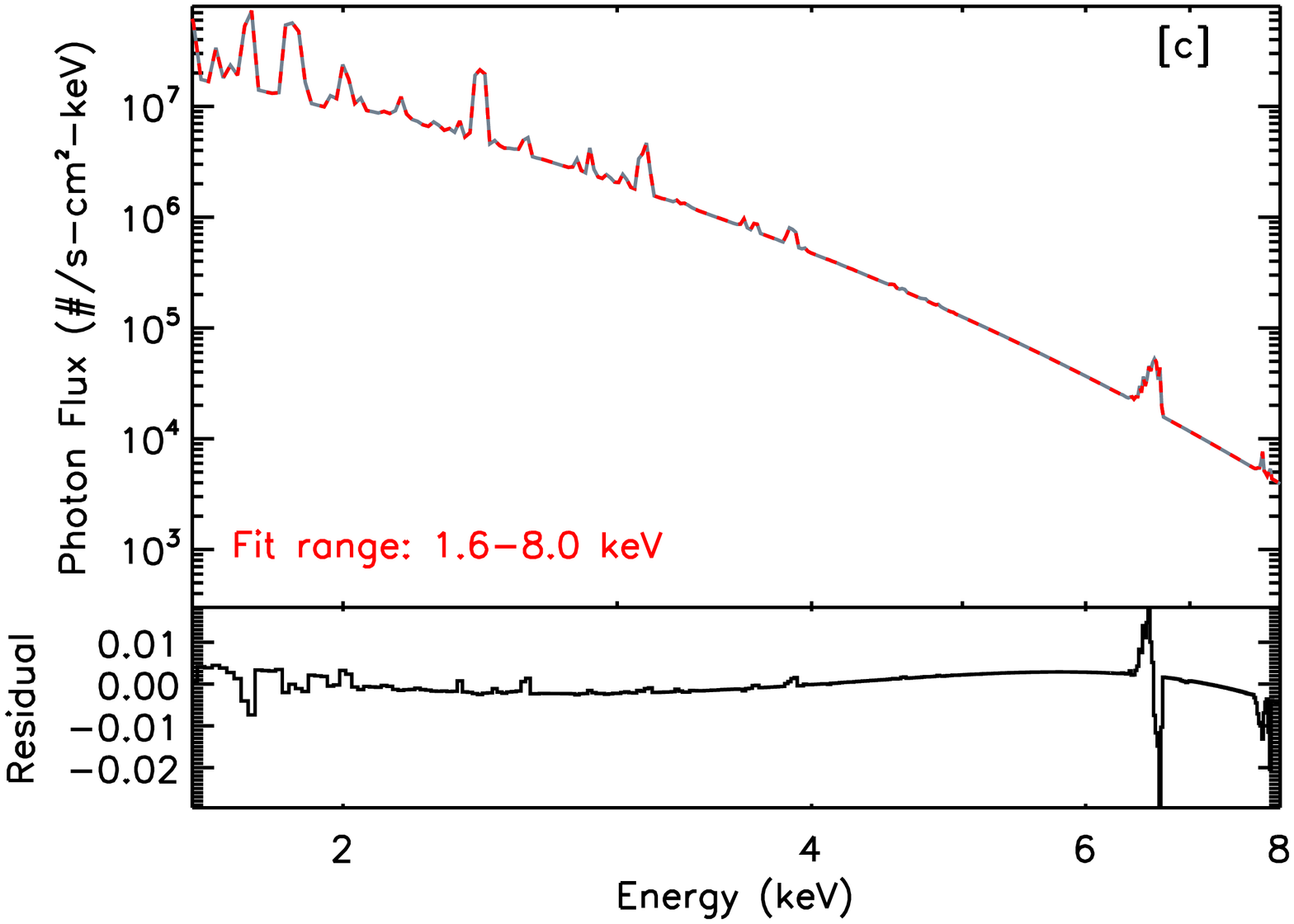, angle=90,  height=0.31\textwidth, clip, trim=0 0 0 0 } &
\epsfig{file=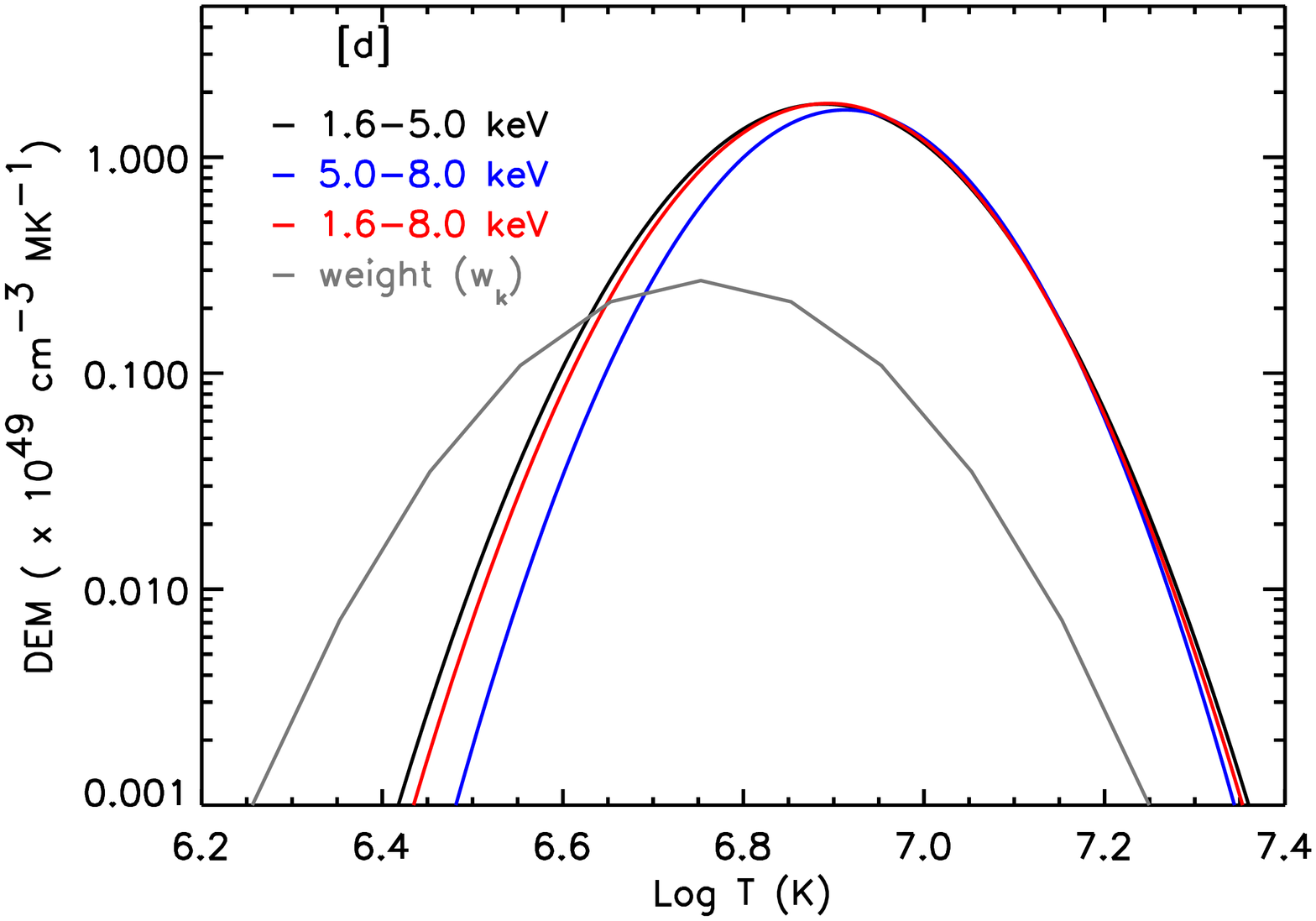, angle=90,  height=0.31\textwidth, trim=0 0 0 0 }
\end{tabular}
\caption{Panels [a], [b] and [c] correspond to model photon spectrum (grey) overlaid by best-fit model flux for 1.6-5.0 keV, 5.0-8.0 keV and 1.6-8.0 keV energy bands, respectively. Normalized residuals are also plotted in the bottom row of all the respective panels. The disagreement of the best-fit curves to the higher and lower part of the spectrum may be noted. Panel [d] presents the DEM[T] corresponding to the best-fit photon flux for all the three cases of energy bands. Normalized weight ($w_k$), employed in Equation \ref{eq-F-multi-thermal}, is also shown by grey color plot. Dotted lines in panels [a] and [b] at 5.0 keV represent the boundary of energy range considered for spectral fit.}
\label{fit-theory-spectra}
\end{figure*}

We synthesize multi-thermal photon spectra by the model photon flux arrays corresponding to iso-thermal plasma in the temperature range of 1-23 MK with a temperature bin of $log~T$=0.1 MK. Iso-thermal photon spectrum at specific temperature and emission measure is calculated by using the iso-thermal model (f\_vth.pro) available in SPectral EXecutive (SPEX) package within \textit{SolarSoftWare (SSW)}. We derive EM values corresponding to a temperature from the emission measure model of \citet{Dere1979}, and also available in CHIANTI atomic database \citep{Landi2012, Del Zanna2015}. In addition we consider the abundance to be 0.1 times the coronal abundance available in the CHIANTI distribution. Next, the multi-thermal photon spectrum is derived by the weighted sum of iso-thermal spectra in following manner.

\begin{equation}
\label{eq-F-multi-thermal}
 F_{MT}=\sum_{k=T_{min}}^{T_{max}} w_{k} F(T_k,EM_k)
\end{equation}
Here $F(T_k,EM_k)$ is the iso-thermal photon flux and shown by grey colour (dotted) plots in Figure \ref{theory-multithermal-spectra} while the multi-thermal flux ($F_{MT}$) synthesized in such a way is over-plotted with red color. Further, $w_k$ is the weight factor which is assumed to be a normalized gaussian function of temperature with the maximum at $T$=5.6 MK and FWHM of $\sim$5 MK as shown in panel [d] of the Figure \ref{fit-theory-spectra}. Such integration allows more realistic scheme of synthesizing theoretical multi-thermal spectra than the one adopted in \citet{Aschwanden2007} and \citet{Jain2011b}. They use direct sum of iso-thermal fluxes with equal weight, in which the synthesized multi-thermal photon spectrum is dominated by the contribution from iso-thermal spectrum corresponding to the peak temperature.

Next, we forward fit the synthesized multi-thermal spectrum using DEM varying as a single-gaussian function of T. The form of the DEM[T] is considered as

\begin{equation}
\label{eq-dem-gauss}
  DEM[T] \propto exp(\frac{-(log T_{p}- log T )^2 }{2\sigma^2})
\end{equation}
where $T_{p}$  is the temperature at the peak of DEM [T] while $\sigma$ is the gaussian width. By iteratively varying the independent variables of Equation \ref{eq-dem-gauss} viz.  $T_{p}$, $\sigma$ etc., a photon flux best-fit to the input synthesized multi-thermal photon spectra is derived. It is to be noted that the minimum and maximum temperature values for deriving the DEM[T] are fixed to 0.087 keV (1 MK) and 8.5 keV (100 MK), respectively. The best-fit is assessed by estimating reduced $\chi^2$ in each step of iteration, which converges to a small value. Following the aforesaid procedure, we derive the best-fit parameters for the synthesized multi-thermal photon spectrum in the low- and high-energy bands. Next, we apply the similar procedure to derive the best fit parameters for the input photon spectrum in the combined-energy band. Panels [a], [b] and [c] of Figure \ref{fit-theory-spectra} show the synthesized input multi-thermal photon spectrum (grey color) overlaid by best-fit model flux drawn in black, blue and red colors for low-, high- and combined-energy bands, respectively. In addition, normalized residuals are also plotted with the respective panels. Moreover, panel [d] presents the DEM evolution corresponding to the best-fit model photon spectra.

From the aforesaid analysis, presented in Figure \ref{fit-theory-spectra}, we find that the best-fit flux derived for the input photon spectrum in low-energy band (panel [a]) does not provide a good fit (over-estimation) to the higher energy part of the spectrum. On the other hand, the best-fit flux obtained for the high-energy band (panel [b]) does not completely follow (under-estimates) the low-energy part of the spectrum. This trend is clearly represented in the form of normalized residuals, plotted with the respective panels. Panel [d] of the Figure \ref{fit-theory-spectra} shows the DEM[T] for the best-fit model photon flux estimated for input photon spectrum corresponding to the low, high and combined energy ranges. This enable us to make a comparative study of the DEM[T] dependence on input energy bands selection. We note that the best-fit DEM[T] curves for the low and high-energy bands yield high peak values of DEM ($DEM_p$)=1.72 $\times$ $10^{49}$ $cm^{-3} MK^{-1}$, however at low temperature $T_p$=7.8 MK for the former (low-energy band) case while relatively lower $DEM_p$=1.66 $\times$ $10^{49}$ $cm^{-3} MK^{-1}$ at higher $T_p$=8.2 MK for the latter case. Moreover, moderate $DEM_p$ at $T_p$ best fits the input spectra of combined-energy band.

\subsubsection{DEM[T] distribution derived from observed X-ray emission during the flare}
We analyze X-ray emission in the low (1.6-5.0 keV) and high (5.0-8.0 keV) energy bands, obtained from SphinX and SOXS, respectively during the flare. In this regard, we prepare a time-series of the spectra by integrating the observed X-ray emission into 120 second time intervals during 04:27-04:33 UT and 04:38-05:00 UT, corresponding to the rise and decay phase of the flare, respectively. On the other hand, better count statistics during 04:33-04:38 UT, corresponding to the impulsive phase of the flare enabled us to integrate the observation in 60 second time intervals. Time sequence of the spectra obtained in such a way serves as the input to the inversion scheme.

\begin{figure*}[!htbp]
\begin{tabular}{cc}
\epsfig{file=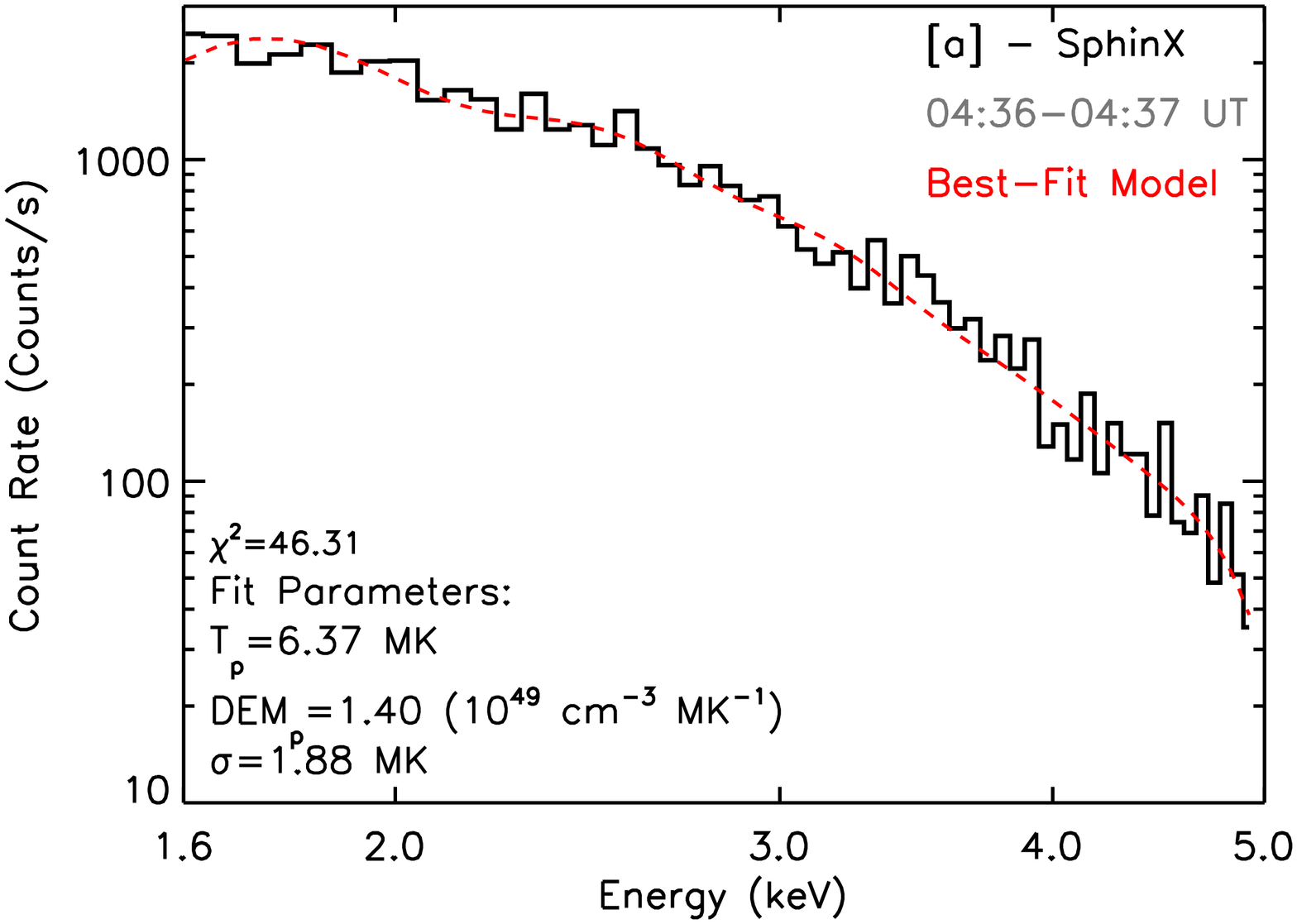, angle=90, height=0.31\textwidth, clip, trim=0 0 0 0} &
\epsfig{file=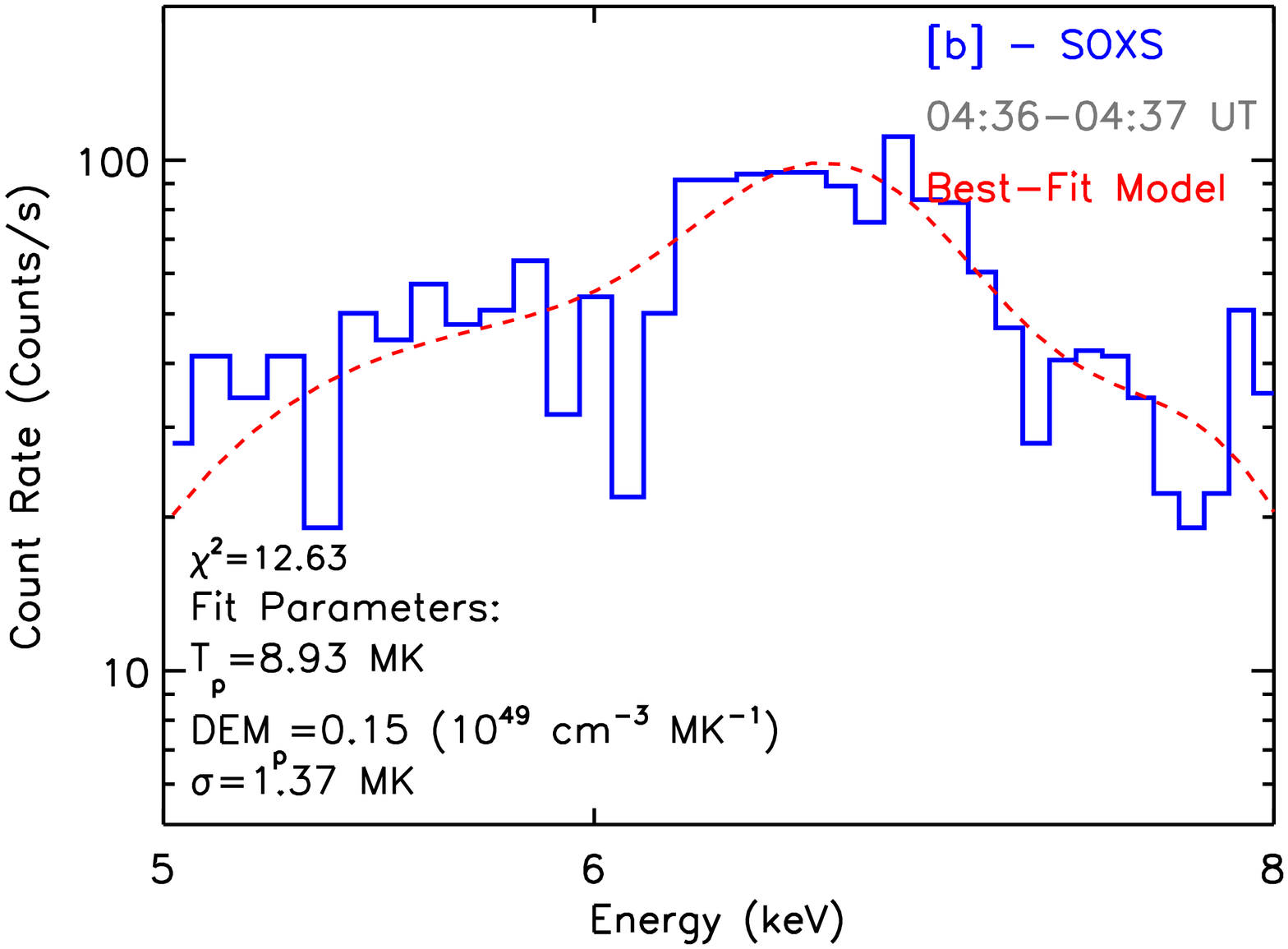, angle=90, height=0.31\textwidth, clip, trim= 0 0 0 0} \\
\epsfig{file=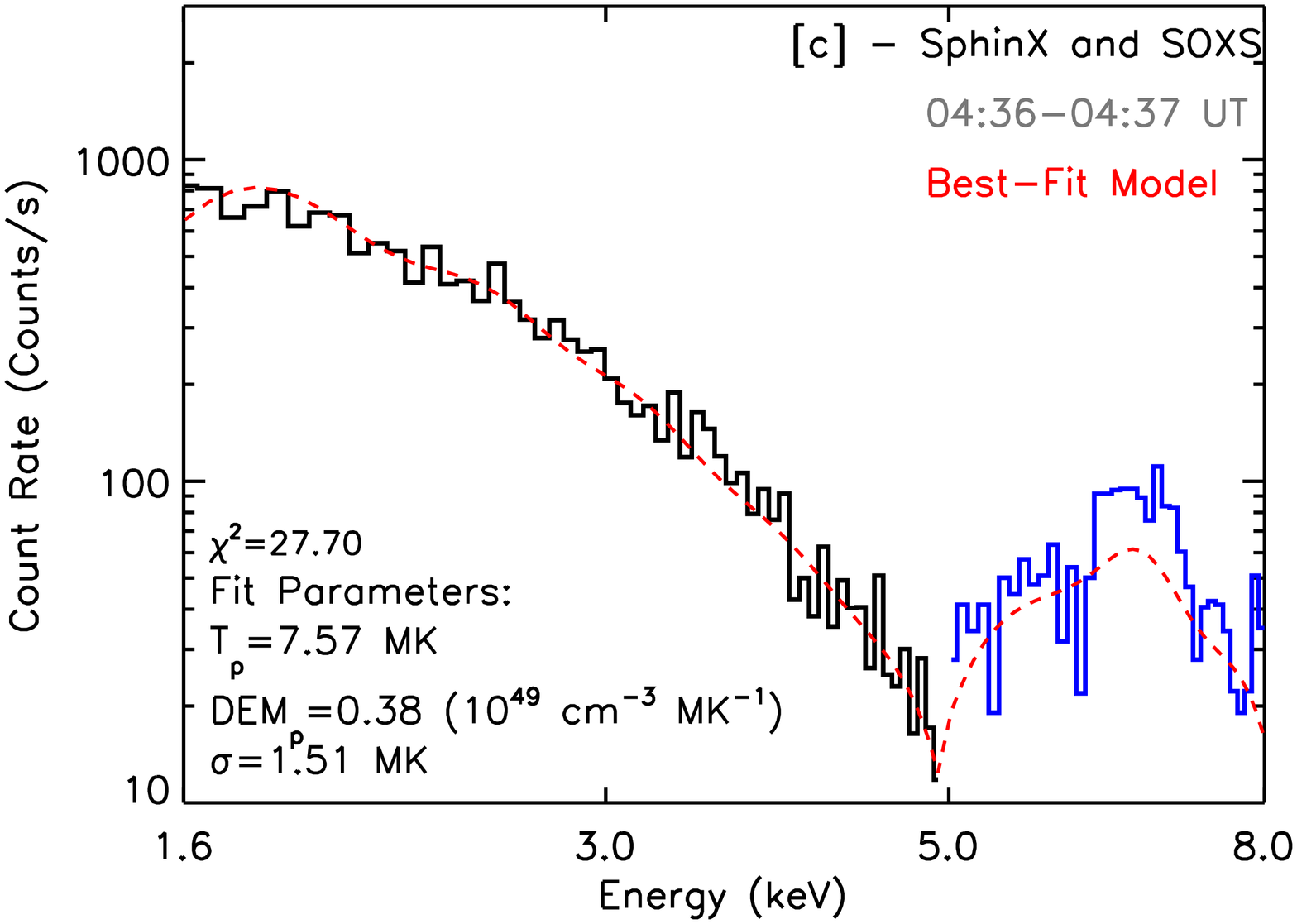, angle=90,  height=0.31\textwidth, clip, trim=0 0 0 0 } &
\epsfig{file=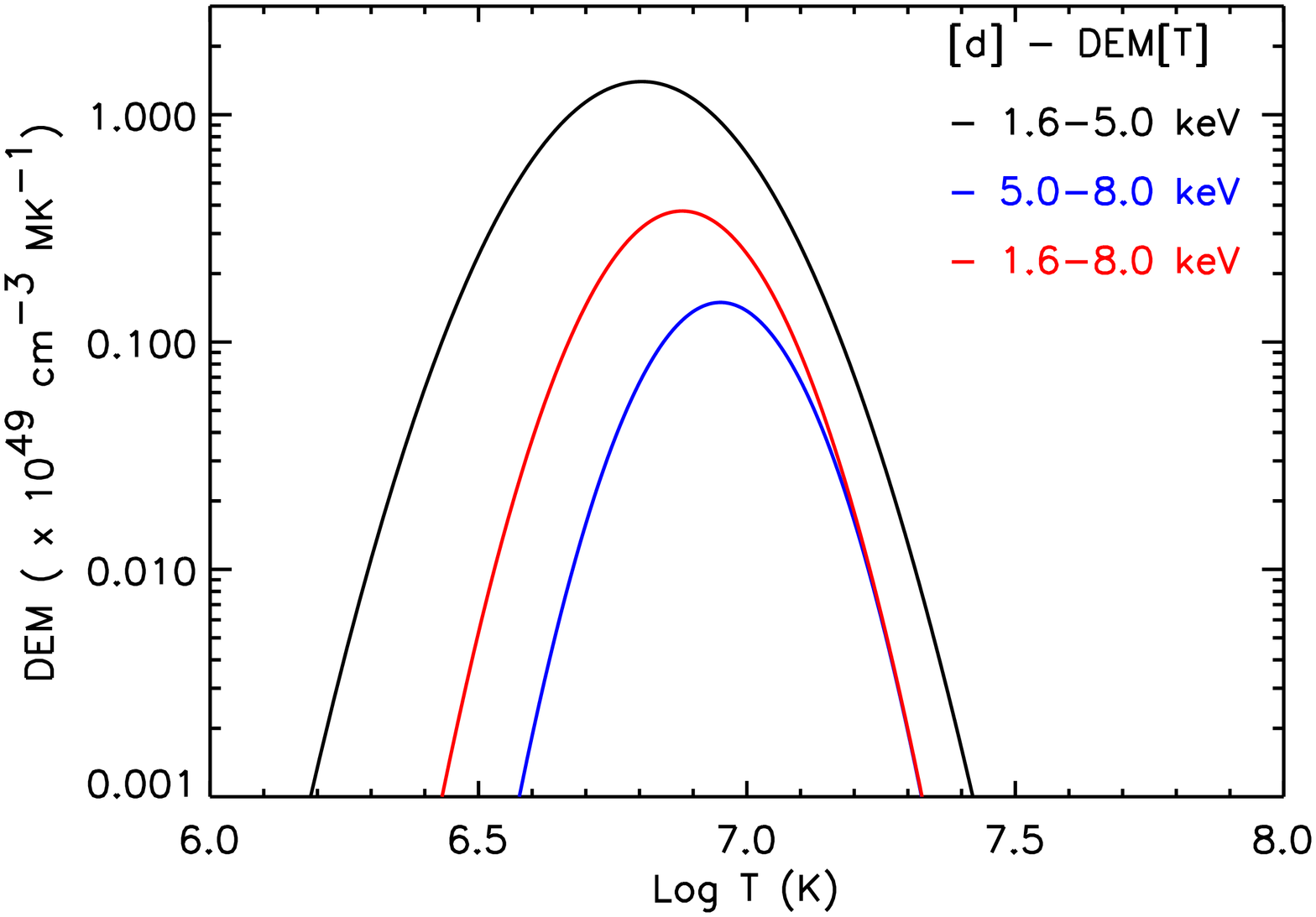, angle=90,  height=0.31\textwidth, trim=0 0 0 0 }
\end{tabular}
\caption{Panels [a], [b] and [c] present the observed spectrum integrated during the maximum phase from SphinX (black) and SOXS (blue) as well as combined data, respectively. Respective best-fit model for the single-gaussian approach is over-plotted by red color. Panel [d] presents the derived DEM[T] curves corresponding to the best-fit model for different energy bands.}
\label{fit-dem-gauss-spectra}
\end{figure*}

We forward-fit the observed SXR spectrum in the low-, high- and combined-energy bands with a model photon flux, which is derived by the inversion scheme employing a single-gaussian functional dependence of the DEM on $T$ (cf. Equation \ref{eq-dem-gauss}). Panel [a] of the Figure \ref{fit-dem-gauss-spectra} presents the observed count rate in low-energy band (black color) during 04:36-04:37 UT, corresponding to the maximum of impulsive phase of the flare. Similarly, X-ray emission in high-energy band (blue color) is analyzed, as shown in panel [b] of the Figure \ref{fit-dem-gauss-spectra} for the aforesaid time interval. Panel [c] presents the analysis of combined-energy band data (from SphinX and SOXS), for the time intervals same as that of panels [a] and [b]. Best fit model count rates are over-plotted by red color in the respective panels and the derived values are also shown. Panel [d] presents the DEM[T] distribution corresponding to the best-fit model obtained for SXR spectra of different energy bands.

From Figure \ref{fit-dem-gauss-spectra}, it may be noted that $DEM_{p}$  estimated from spectral fitting of the low, high and combined-energy band data are 1.40, 0.15 and 0.38 ($ \times 10^{49} \mathrm{cm^{-3} MK^{-1}}$), respectively. On the other hand, $T_p$ is estimated to be 6.37, 8.93 and 7.57 MK, respectively. It may be noted that the trend of the best-fit parameters viz. $DEM_{p}$ and $T_p$ for three cases of input energy band is in good agreement with that revealed by the study of model multi-thermal spectrum employing the same inversion scheme as presented in the previous section.

\subsection{DEM varying as power-law function of T}
We derive the DEM[T] distribution for the X-ray spectra corresponding to various energy bands, similar to the analysis made in the previous section, however with different functional dependence of DEM on $T$. In this multi-thermal model, DEM is approximated to be varying with $T$ in the form of power-law and can be expressed as:

\begin{equation}
 DEM(T) \propto (\frac{2}{T})^\gamma
\end{equation}

\begin{figure*}[!htbp]
\begin{tabular}{cc}
\epsfig{file=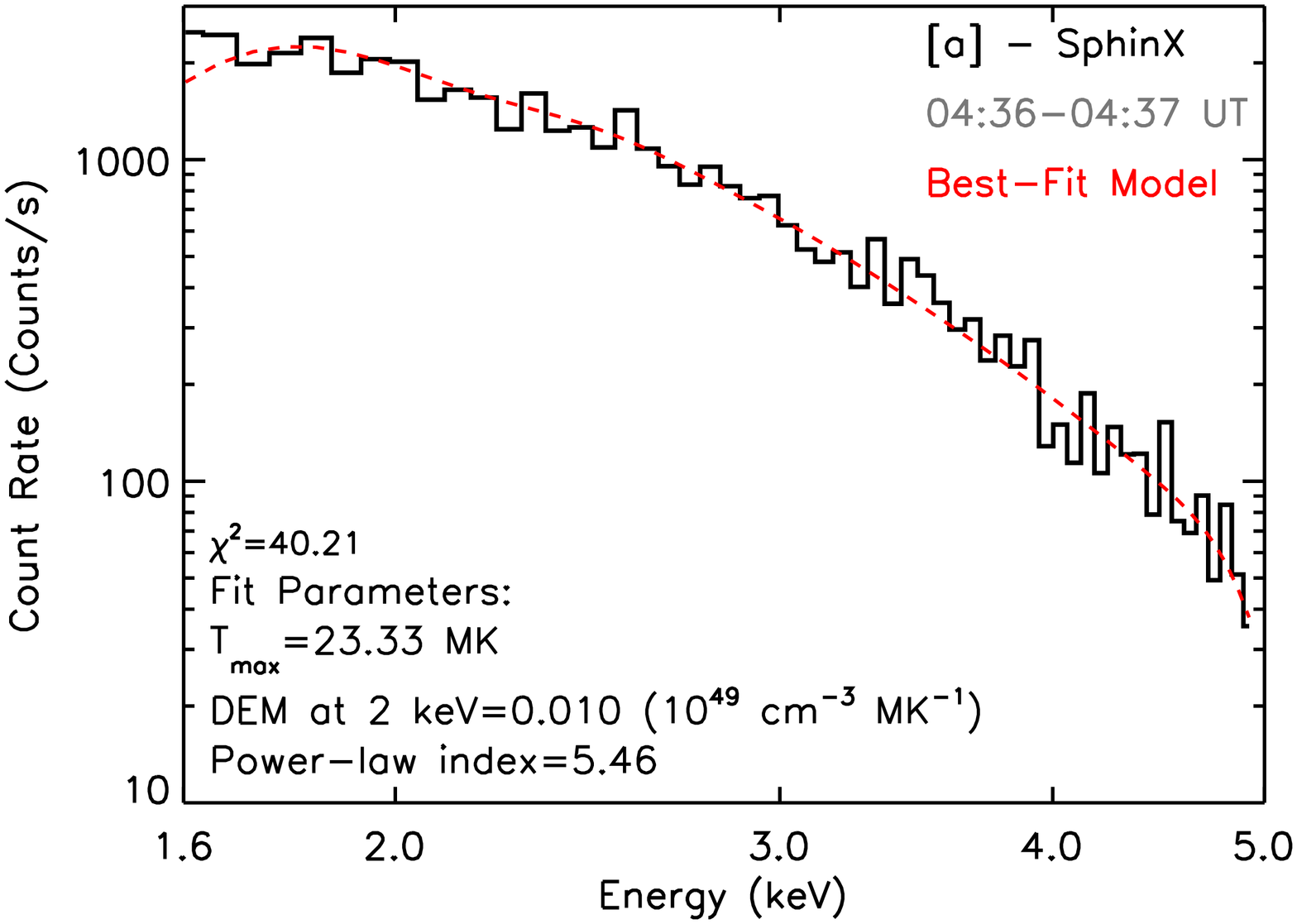, angle=90, height=0.31\textwidth, clip, trim=0 0 0 0} &
\epsfig{file=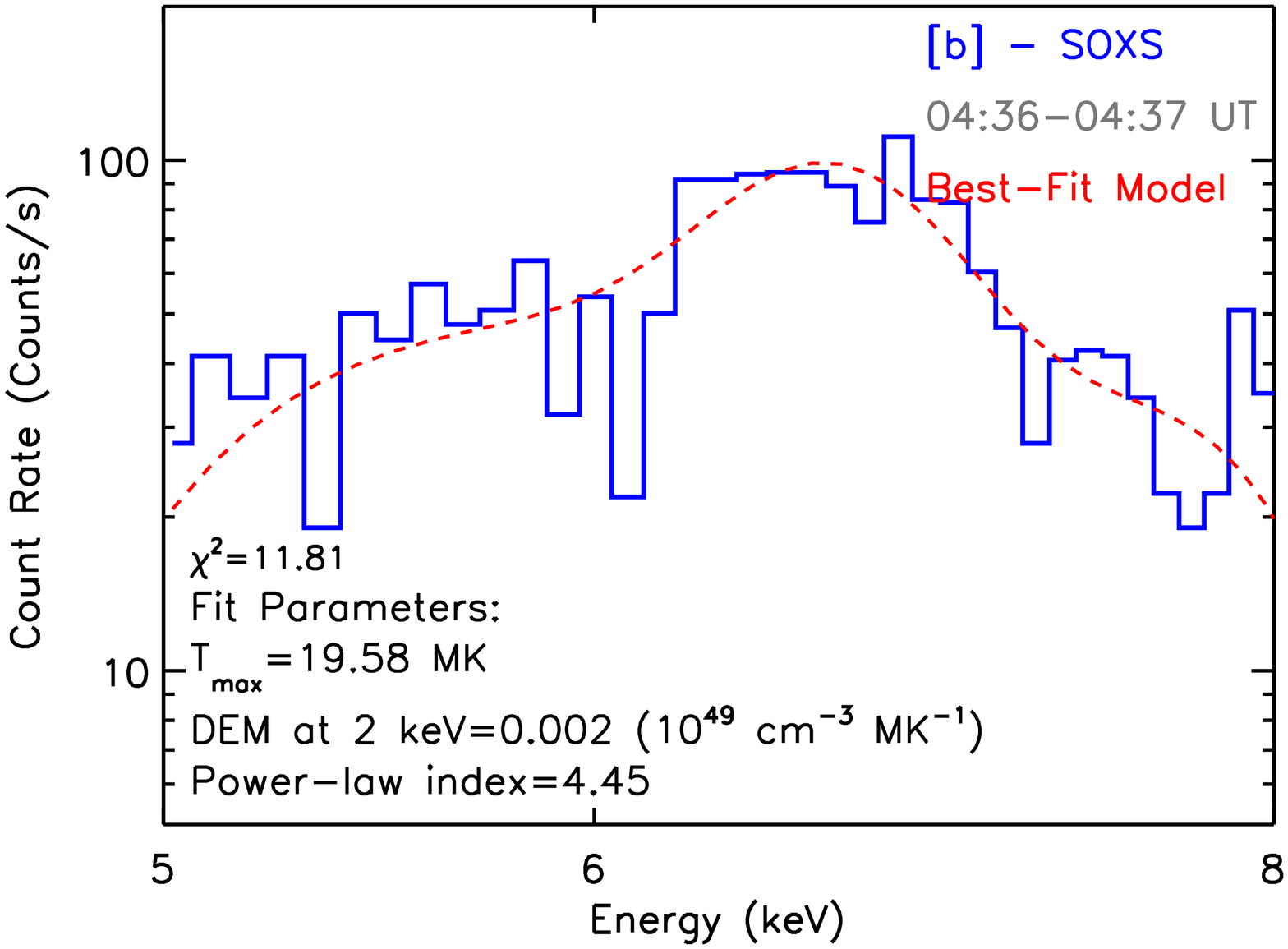, angle=90, height=0.31\textwidth, clip, trim= 0 0 0 0} \\
\epsfig{file=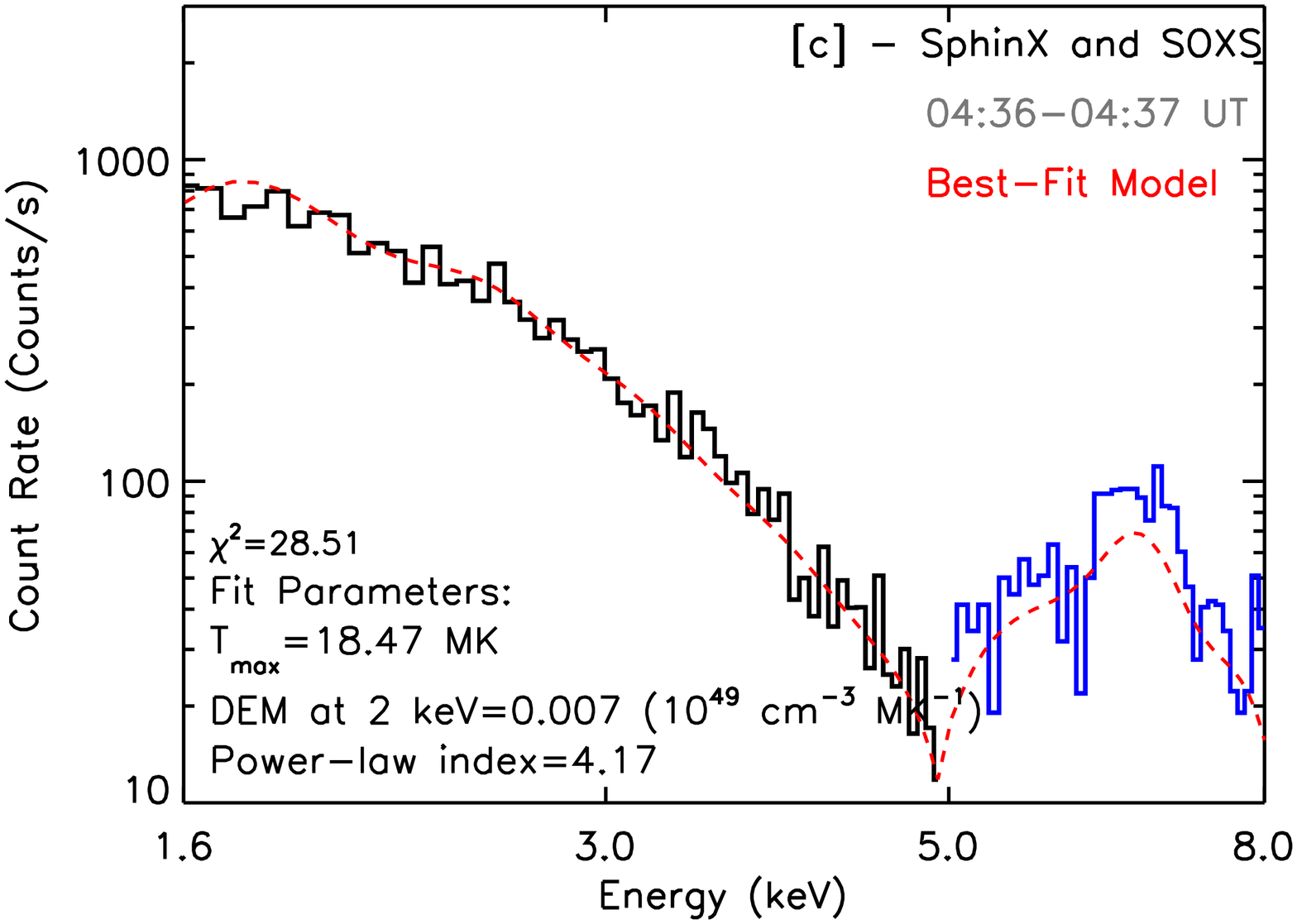, angle=90, height=0.31\textwidth, clip, trim=0 0 0 0} &
\epsfig{file=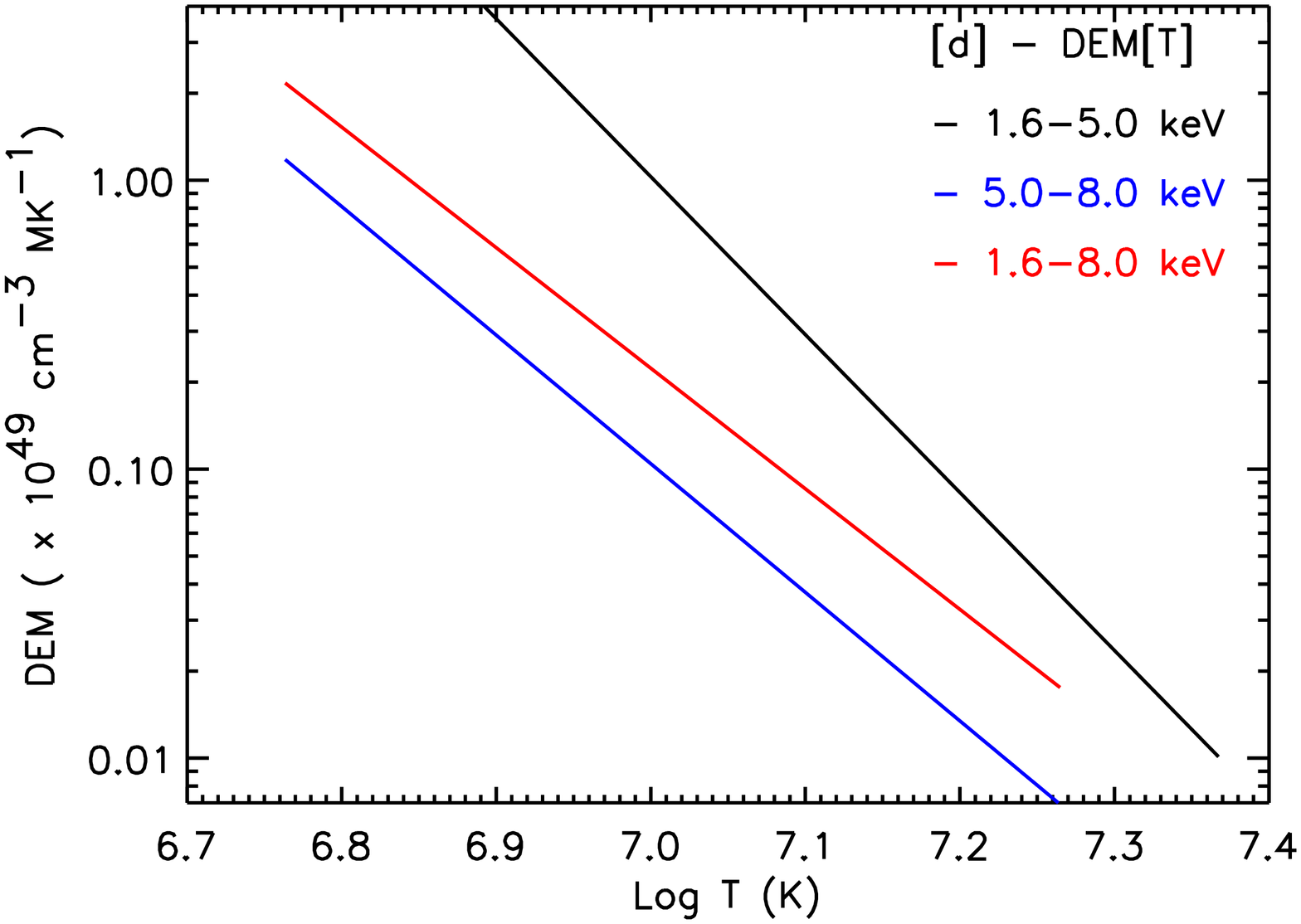, angle=90, height=0.31\textwidth, trim=0 0 0 0}
\end{tabular}
\caption{Panels [a] and [b] show the observed count fluxes measured by SphinX (black) and SOXS (blue) during 04:36-04:37 UT, respectively. Panel [c] shows the combined X-ray spectra in 1.6-8.0 keV for the time interval same as that in panels [a] and [b]. The red curve, over-plotted on respective panels, represents the best-fit model count flux, derived employing the power-law DEM[T] scheme. Panel [d] presents the derived DEM[T] distribution corresponding to best-fit model count flux for different energy ranges.}
\label{fit-dem-power-law-spectra}
\end{figure*}

Next, employing this DEM scheme, we forward-fit the observed flare X-ray spectrum in low-energy band obtained from SphinX (black color) as shown in panel [a] of the Figure \ref{fit-dem-power-law-spectra}. During the iterative procedure for obtaining the best-fit model, the low-temperature value is fixed to be 0.5 keV (5.8 MK) while the maximum temperature is determined as one of the outputs. All the spectra during various time intervals of the flare are analyzed, however here we present only the results from the observations during 04:36-04:37 UT, same as that presented in the previous section. X-ray emission in high-energy band, obtained from SOXS (blue color) is presented in panel [b] of the Figure \ref{fit-dem-power-law-spectra}. Next, we also fit the observed X-ray spectrum in combined-energy band as shown in panel [c]. The best-fit models are over-plotted by red color lines in the respective plots. Parameters of the best-fit viz. $T_{max}$, DEM (at T=2 keV) and the power-law index ($\gamma$) are also shown in respective plots. Panel [d] shows the derived DEM[T] curves corresponding to the best-fit models obtained for different energy ranges.

From Figure \ref{fit-dem-power-law-spectra}, it may be noted that at the peak of the impulsive phase of the flare, $T_{max}$ estimated from the observation recorded by SphinX, SOXS as well as combined observations is 23.33, 19.58 and 18.47 MK, respectively. It may be noted that $T_{max}$ and DEM values estimated in such a way follow the same trend as that resulted in the previous DEM scheme. Moreover, the negative power-law index ($\gamma$) of the best-fit DEM[T] distribution corresponding to the SXR spectrum in low, high and combined-energy bands is estimated to be 5.46, 4.45 and 4.17, respectively. The less-negative (steeper) value of `$\gamma$' for the high and combined-energy cases suggest enhanced contribution of high-temperature plasma than that obtained from the analysis of SXR emission in the low-energy band only.

\subsection{Withbroe-Sylwester (W-S) maximum likelihood DEM inversion algorithm}
We employ Withbroe-Sylwester (W-S) maximum likelihood DEM inversion algorithm \citep{Sylwester1980, Kepa2006, Kepa2008} on the X-ray spectra observed during the flare. The W-S algorithm is a Bayesian numerical technique which employs maximum likelihood approach in which the DEM distribution in one step of iteration `j' ($DEM_{j}[T]$) is estimated from that derived in the preceding iteration ($DEM_{j-1}[T]$), and by employing a correction factor ($c_i$) as well as weight factor ($w_i$) in the form given below.

\begin{equation}
 DEM_{j}[T]=DEM_{j-1}[T] \frac{\sum_{i=1}^{k}c_i w_i(T)}{\sum_{i=1}^{k}w_i(T)}
\end{equation}
Here, the correction factor, $c_i$, is estimated from the ratio of the observed flux with the calculated flux, which is derived using previous DEM distribution form and can be expressed mathematically as:

\begin{equation}
c_{i}=\frac{F_{obs, i}}{F_{cal, i}}
\end{equation}
where, $F_{cal,i}$ is the calculated model flux and obtained by,

\begin{equation}
F_{cal,i}=\int_{j=0}^ \infty f_i(T) DEM_j(T) dT
\end{equation}
In the aforesaid function, $f_i(T)$ is the theoretical emission function for energy `i' and is derived using the CHIANTI package \citep{Del Zanna2015}. The weight factor $w_i$ is estimated as:

\begin{equation}
\begin{split}
w_i(T)=f_i(T) DEM_j(T) dT \frac{\int_{j=0}^ \infty f_i(T) DEM_j(T) dT}{\int_{j=0}^ \infty [f_i(T) DEM_j(T)]^2 dT} \times \\
[\frac{|F_{obs,i}-F_{cal,i}|}{\delta_i}+1]^a
\end{split}
\end{equation}
Here, $\delta_i$ is the uncertainty corresponding to the observations for energy `i' and `a' is termed as the speed convergence parameter.

\begin{figure*}[!htbp]
\begin{center}
\begin{tabular}{c}
\vspace{1em}
\epsfig{file=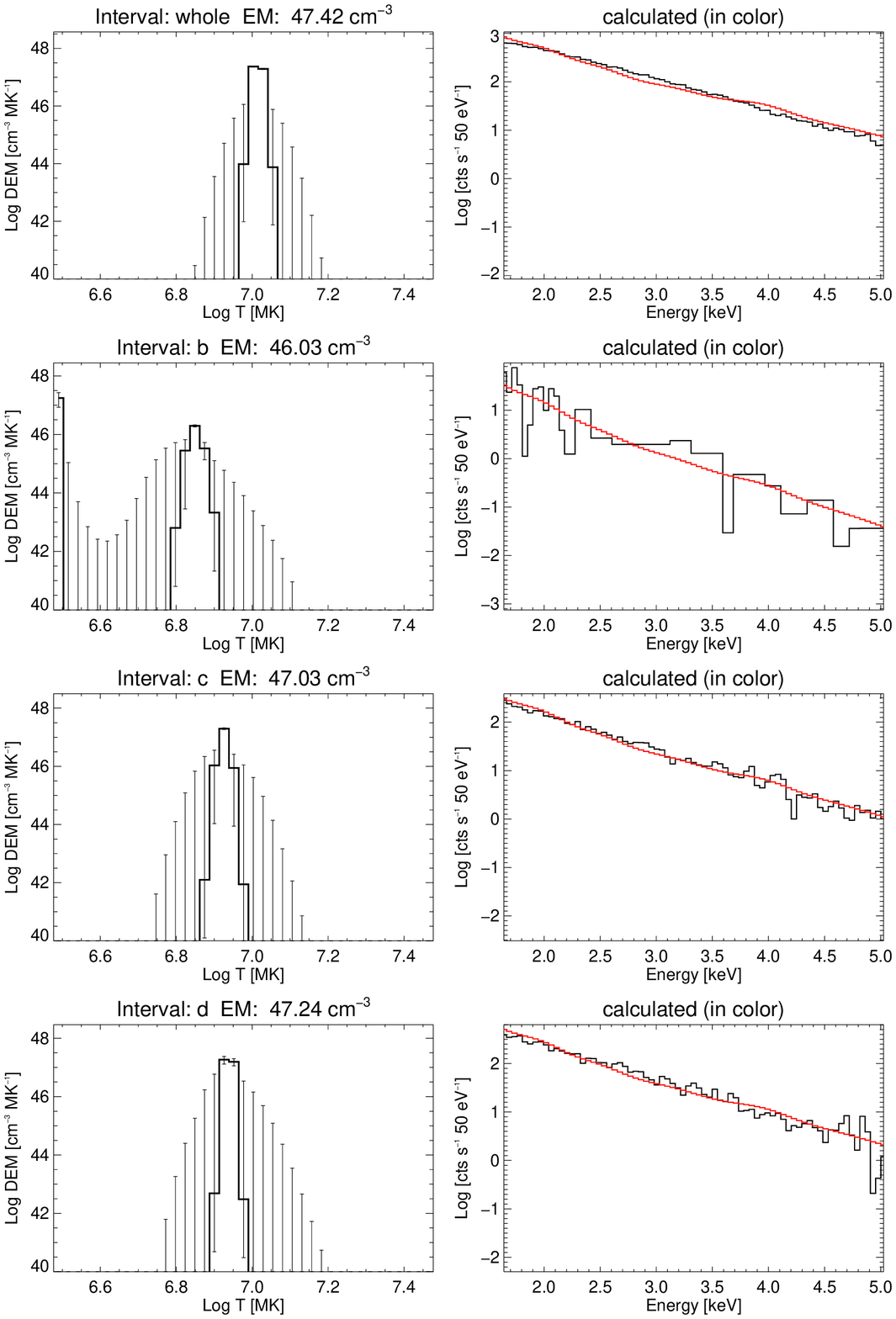, angle=0, width=0.8\textwidth,  clip, trim=0 520 0 28} \\
\epsfig{file=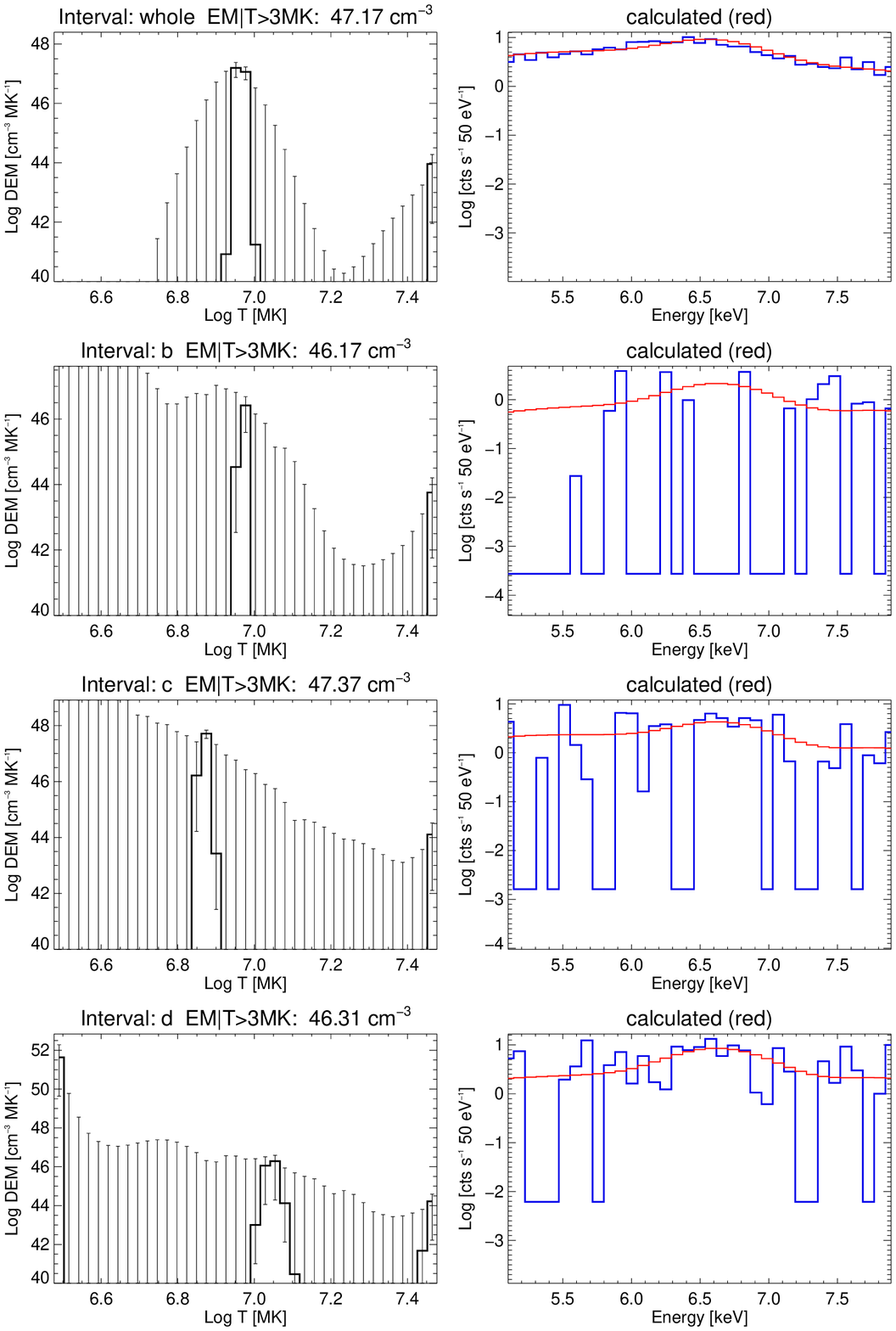, angle=0, width=0.8\textwidth,  clip,  trim= 0 520 0 28} \\
\epsfig{file=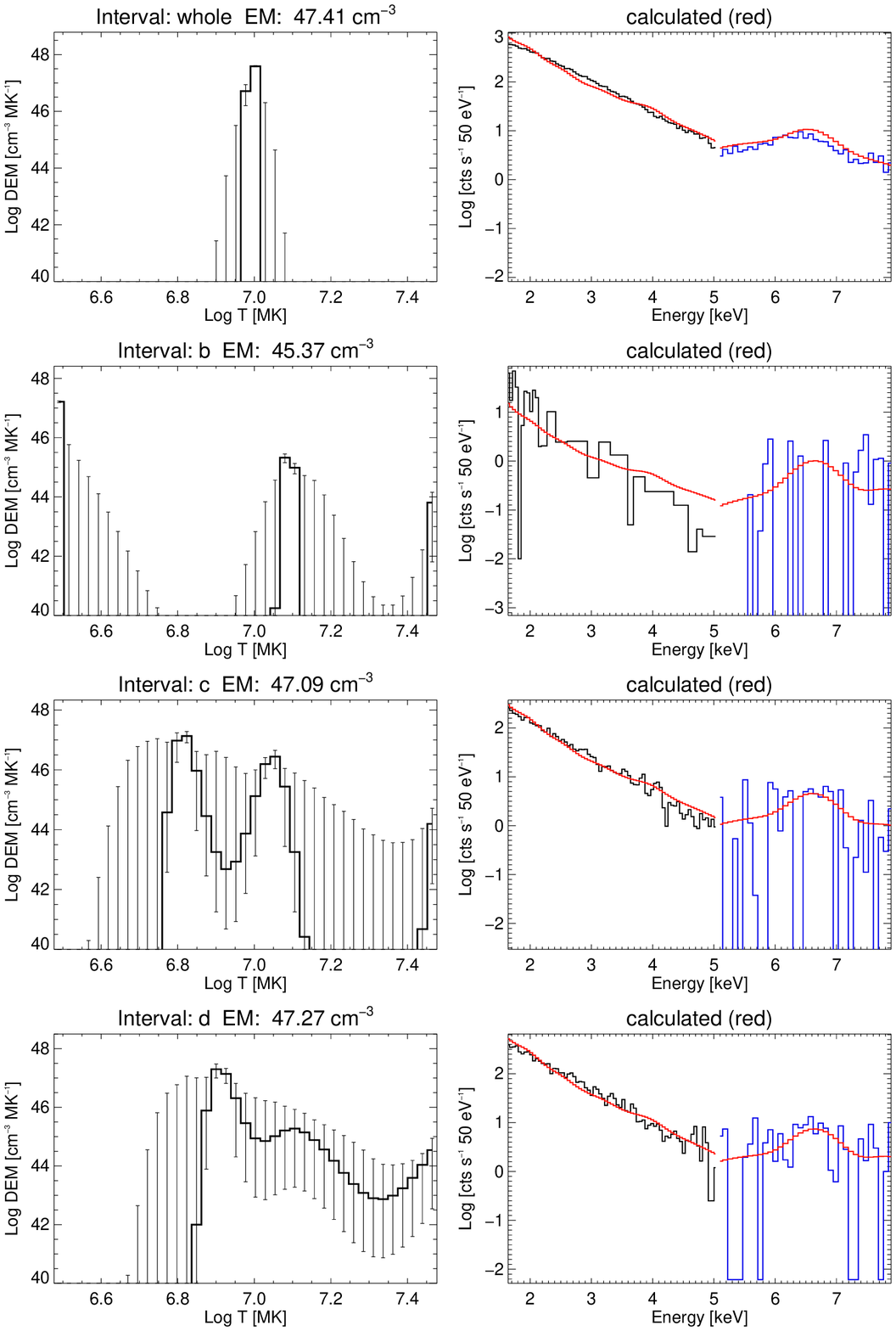, angle=0, width=0.8\textwidth,  clip, trim= 0 520 0 28} \\
\end{tabular}
\caption{Left column presents the DEM[T] curves related to the best-fit model flux (red), which are derived by employing W-S procedure on the observed X-ray emission in: 1.6-5.0 keV (black), recorded by SphinX (top), 5.0-8.0 keV (blue) obtained from SOXS (middle) and 1.6-8.0 keV from SphinX \& SOXS (bottom). The observed X-ray spectrum has been integrated for the time range 04:27:30-05:00:00 UT, covering the entire flare duration.}
\label{w-s-dem-fit-total-sphinx-soxs}
\end{center}
\end{figure*}

We apply the aforesaid W-S DEM inversion algorithm on the X-ray spectra obtained from SphinX and SOXS missions during the flare to obtain the best-fit photon flux and corresponding DEM[T] distribution. Coronal abundances from CHIANTI atomic database have been adopted while calculating theoretical dependence of spectral shapes. Top and the middle rows of the Figure \ref{w-s-dem-fit-total-sphinx-soxs} present the results of the application of W-S algorithm on the X-ray emission measured by SphinX, SOXS, respectively, during 04:27:30-05:00:00 UT, covering the entire flare duration. Moreover, bottom panel of the Figure \ref{w-s-dem-fit-total-sphinx-soxs} shows the same, however, corresponding to the combined data-set. Left panel shows the DEM[T] distributions obtained from the best-fit model (red) for the observations, shown in the right column.

From the application of W-S scheme, as shown in Figure \ref{w-s-dem-fit-total-sphinx-soxs}, we find the peak temperature ($T_{p}$)= 10.0, 9.5 and 10.0 MK, and the total emission measure (EM) log(EM)=47.42, 47.17 and 47.41 ($cm^{-3}$), corresponding to the SphinX, SOXS and combined-energy band data, respectively. This suggests that the trend of the parameters obtained by the W-S scheme for the three cases of the input energy bands is in agreement with that obtained from the previous schemes.

\begin{figure*}[!htbp]
\centering
\begin{tabular}{c}
\vspace{1em}
\epsfig{file=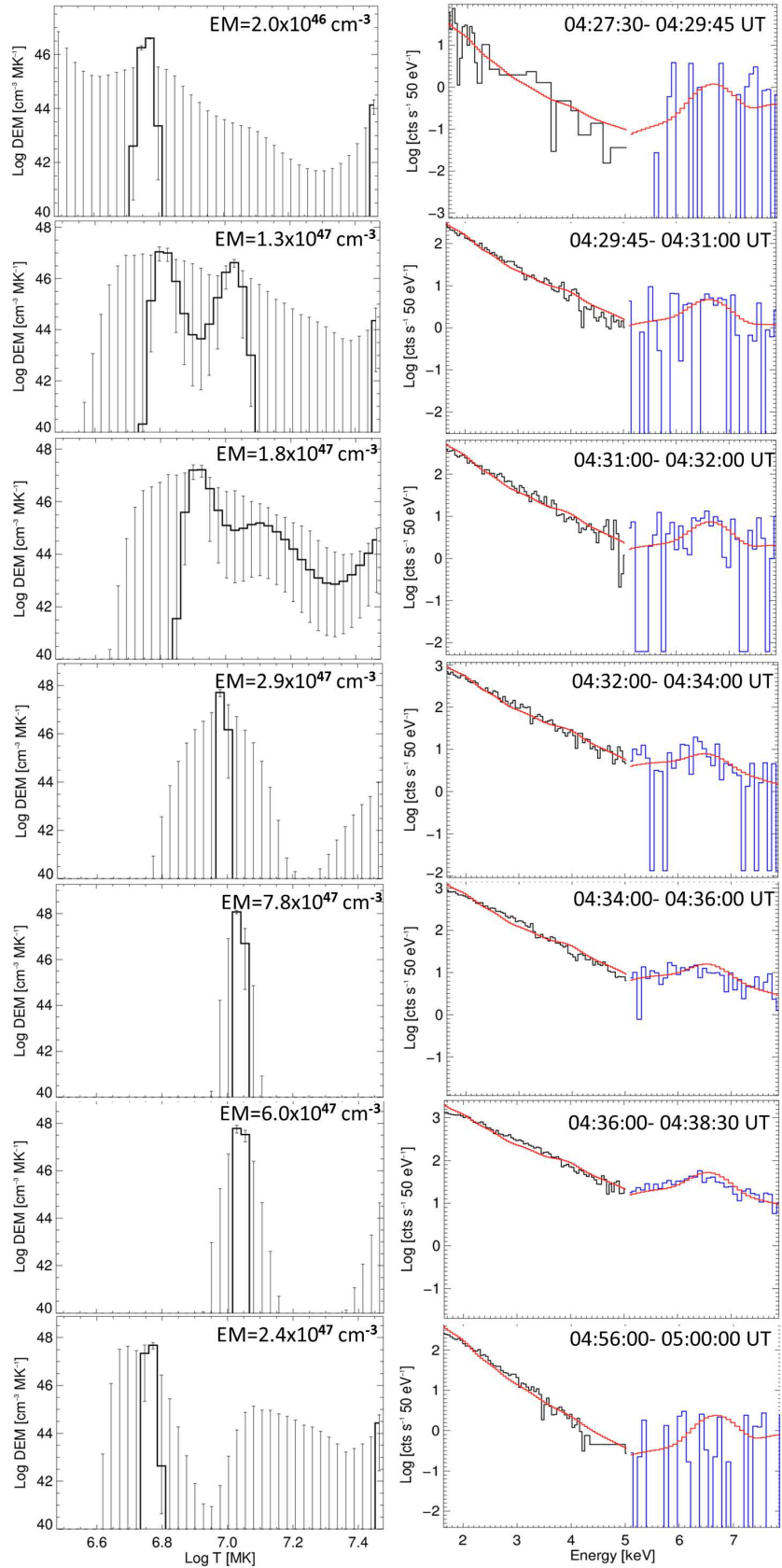, angle=0, height=0.95\textheight,  clip}
\end{tabular}
\caption{Temporal evolution of best-fit DEM[T] distribution (left column) derived using W-S procedure from the SXR emission (right column) recorded in various phases of the flare. Observed X-ray spectra in 1.6-8.0 keV (1.6-5.0 and 5.0-8.0 keV, observed by SphinX and SOXS are plotted with black and blue colors, respectively) and corresponding best-fit model is overlaid in red color.}
\label{w-s-dem-fit-sphinx-soxs}
\end{figure*}

The aforesaid analysis is made for the spectra obtained by integrating the emission in the whole flare duration. Next, we derive the temporal evolution of DEM[T] distribution during various phases of the flare by applying W-S algorithm on the X-ray emission observed during various time intervals of the flare, as presented in Figure \ref{w-s-dem-fit-sphinx-soxs}. Left panels of the Figure \ref{w-s-dem-fit-sphinx-soxs} show the temporal evolution of best-fit DEM[T] distribution derived over various time intervals of the flare, while the respective right panels show the observed X-ray spectra in the combined-energy band (1.6-5.0 and 5.0-8.0 keV, observed by SphinX and SOXS) overlaid by best-fit model (red).

From Figure \ref{w-s-dem-fit-sphinx-soxs}, it may be noted that the best fit DEM[T] curve, obtained from the analysis of the X-ray emission measured during the flare onset time (04:27:30-04:29:45 UT), can be well approximated by a single gaussian function of T with a width $\sim$1 MK. Moreover, the peak temperature ($T_p$) is estimated to be 5.62 MK. On the contrary, the best-fit DEM[T] curve obtained by analysing the spectra during 04:31-04:34 UT, corresponding to the rise phase of the flare, resembles the double-peak gaussian with the increased widths (in comparison to that during the flare onset) $\sim$1.5 MK . Moreover, $T_p$ is estimated to be varying in the range of 6.3-14.1 MK. This reveals the signature of contribution of high temperature plasma in this phase in addition to the low-temperature component, which was present during the flare onset. Further, DEM[T] derived for the spectra obtained during 04:32-05:00 UT, corresponding to the peak of the impulsive phase and decay phase of the flare, resulted in single peak gaussian nature, however, with peak temperature varying in the range of $\sim$ 13.0-5.5 MK.

It is intriguing to note that the best-fit DEM[T] distribution, obtained by integrating the emission in the whole flare duration, as shown in Figure \ref{w-s-dem-fit-total-sphinx-soxs}, can be well approximated to iso-thermal nature. On the contrary, temporal evolution of DEM[T] distribution over various phases of the flare suggests the presence of multi-thermal plasma during the rise phase of the flare. This apparent inconsistency may be explained by the fact that if the X-ray spectrum is integrated for the whole flare duration, it is dominated by the emission at the peak of the impulsive phase. Now, it may be noted that the best-fit DEM[T], derived for the spectrum during  04:36:00-04:38:30 UT (corresponding to the peak of the impulsive phase, see Figure \ref{w-s-dem-fit-sphinx-soxs}) is iso-thermal in nature.

\begin{figure*}[!htbp]
\begin{tabular}{cc}
\vspace{1em}
\epsfig{file=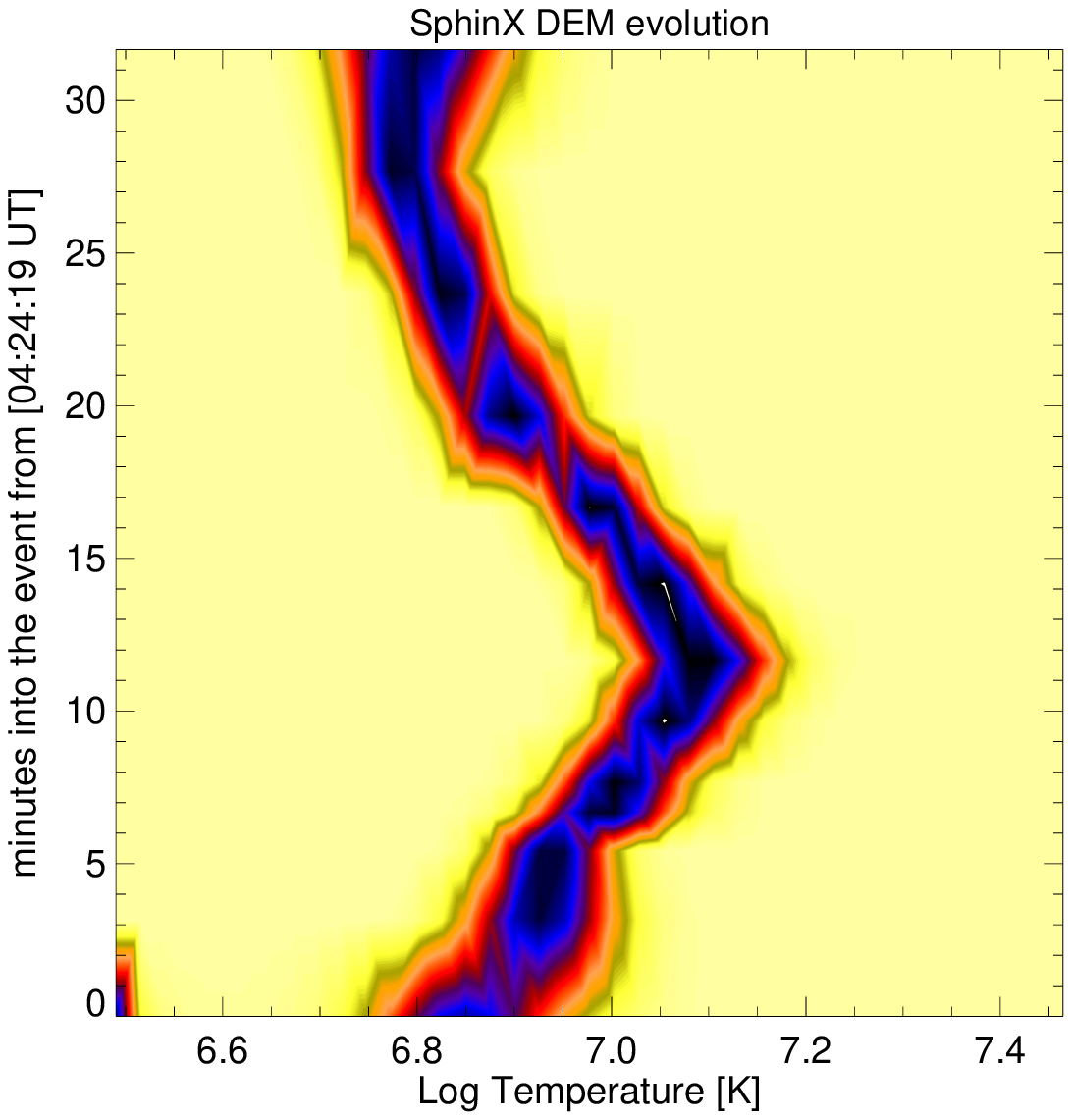, angle=0, width=0.41\textwidth,  clip, trim=0 5 80 160} &
\epsfig{file=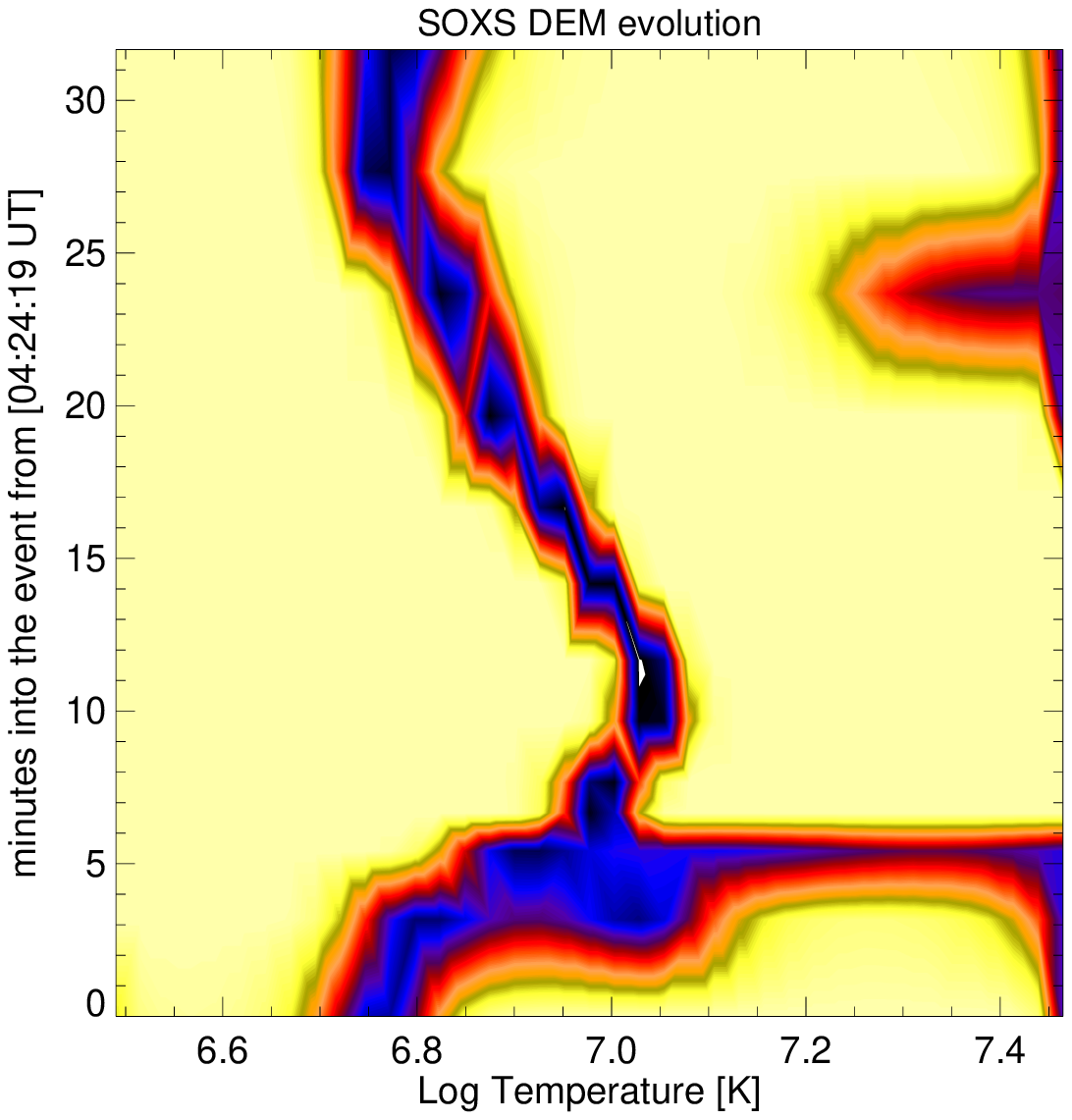, angle=0, width=0.41\textwidth,  clip,  trim= 0 5 80 160}
\end{tabular}
\caption{Temporal evolution of DEM[T] distribution derived from SphinX alone (left) and combined SphinX \& SOXS (right) observations.}
\label{dem-3d-sphinx+combined}
\end{figure*}

In Figure \ref{dem-3d-sphinx+combined}, we present the different visualisation of temporal evolution of DEM[T] distribution over the flare duration, from the SphinX observations in the left panel while that from the combined-energy band observations in the right panel. The comparison of DEM[T] derived from the  SphinX observations alone with that obtained from combined observations reveals the signature of high temperature component in the latter analysis during the rise-phase of the flare.

\section{Thermal energetics of the flare}
\label{sec:e-th}
We estimate thermal energy content during various phases of the flare. We denote the energy content which is estimated employing the iso-thermal approach as the `iso-thermal energy' while that derived considering flare plasma to be of multi-thermal nature is termed as `multi-thermal energy'. Next, we make a comparative study of the multi-thermal energy content derived by the application of various DEM inversion schemes.

\begin{figure*}[!htbp]
\begin{tabular}{ccc}
\epsfig{file=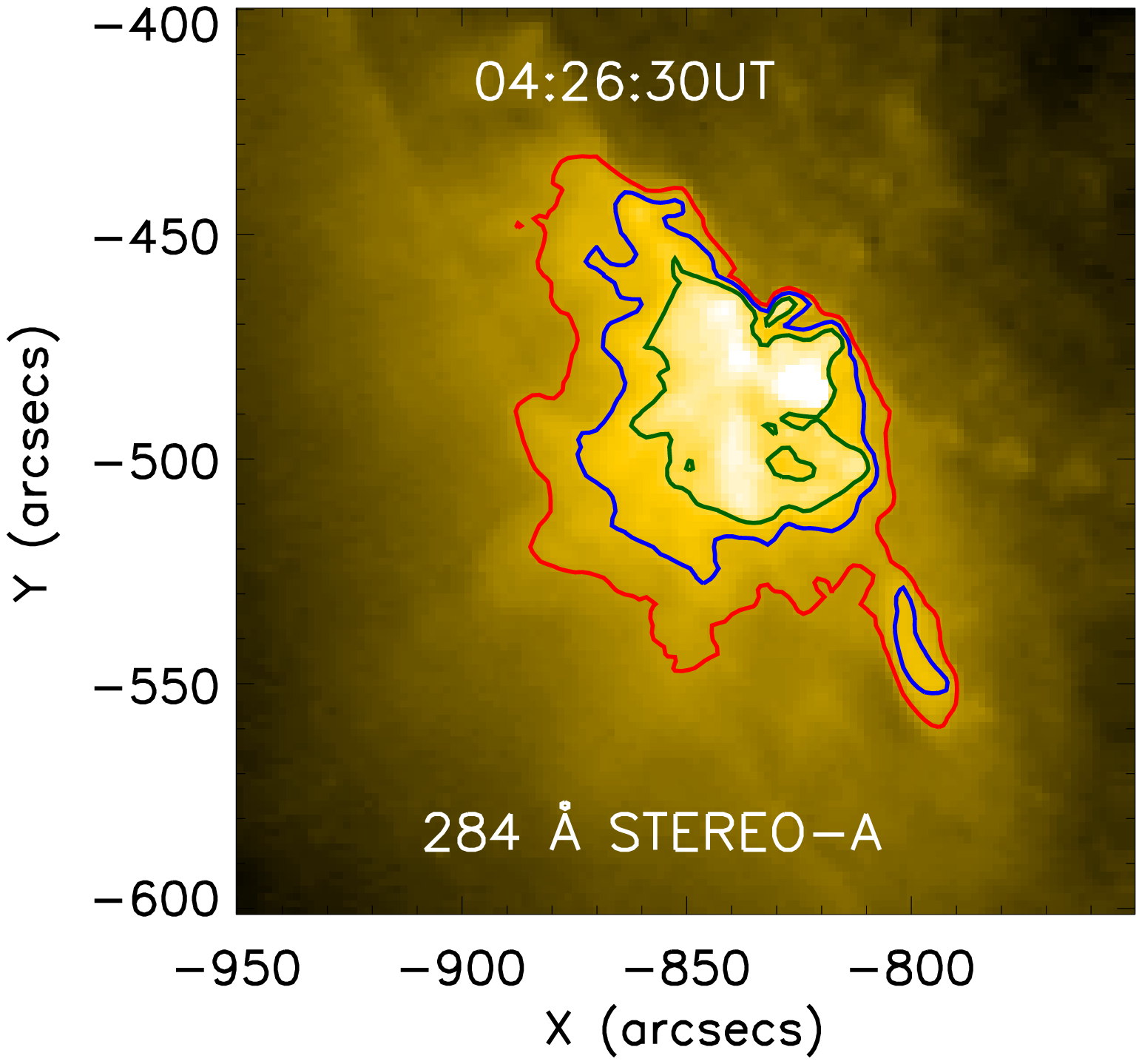, angle=90, height=0.27\textwidth,  clip, trim=20 170 70 70} &
\epsfig{file=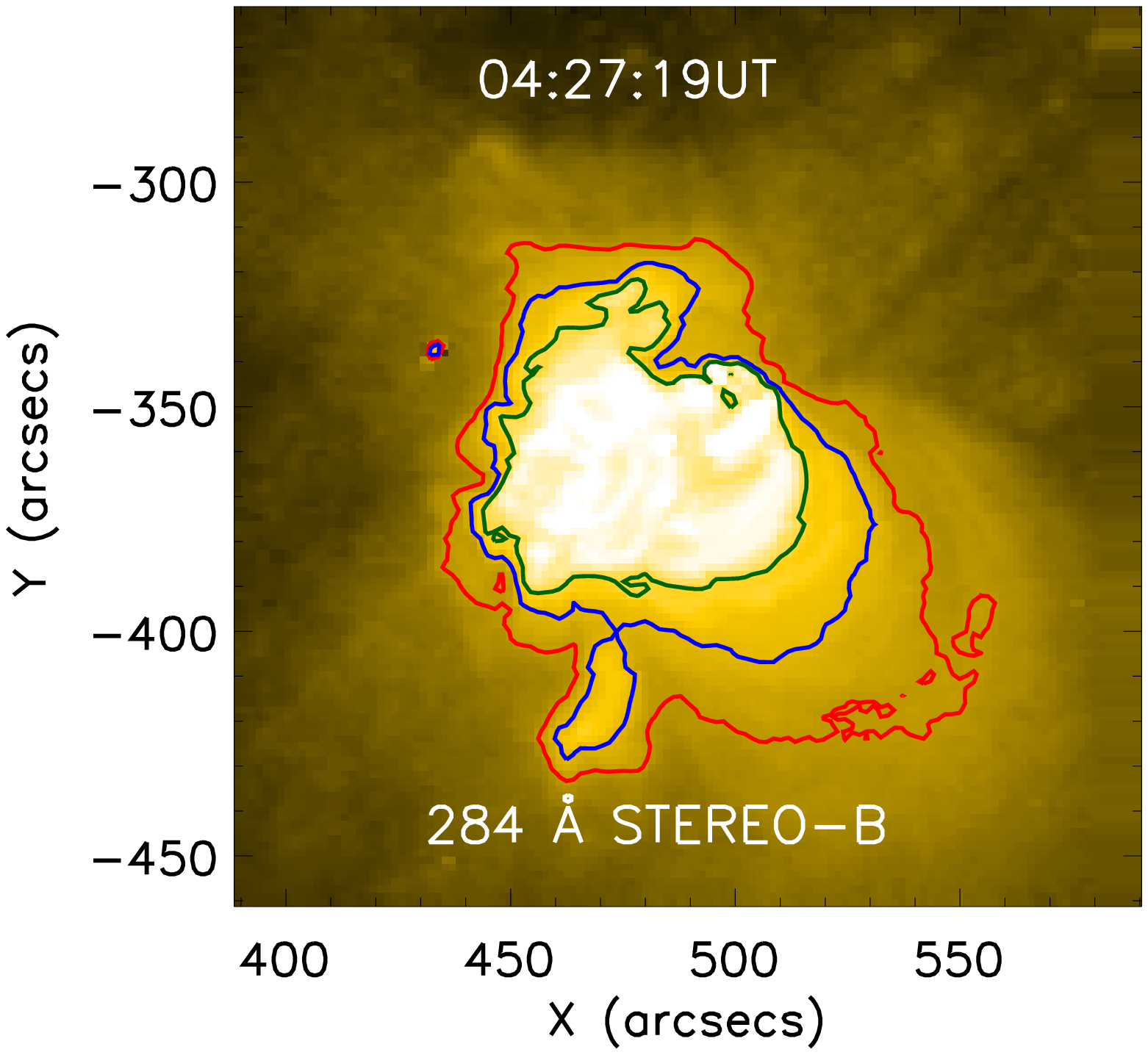, angle=90, height=0.27\textwidth,  clip,  trim= 20 170 70 70} &
\epsfig{file=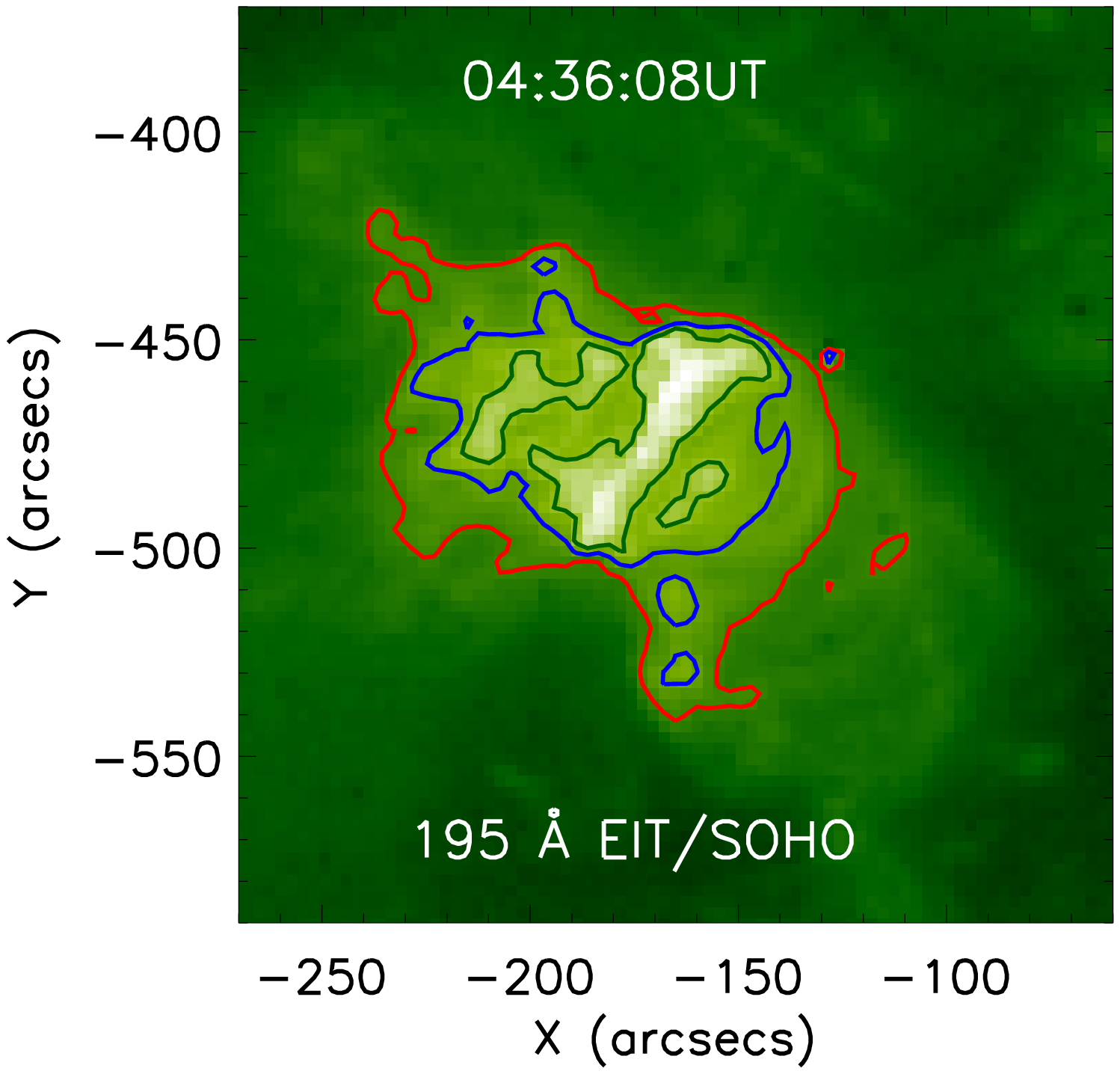, angle=90, height=0.27\textwidth,  clip, trim= 20 170 70 70} \\
\end{tabular}
\caption{Images in 284 $\mathrm{\AA}$ during the flare, obtained by \textit{STEREO-A} and \textit{B} as shown in left and middle panels, respectively. Right panel shows the image obtained during the flare at 195 $\mathrm{\AA}$ from EIT/\textit{SOHO}. The contours overlaid on the images correspond to 5\% (red), 10\% (blue) and 20\% (green) of the maximum intensity regions.}
\label{st-a-b-eit-contour}
\end{figure*}

In order to estimate the iso-thermal energy content of the flare plasma, we derive the temperature ($T$) and emission measure ($EM$) by employing the technique presented in \citet{Gburek2013} on the high temporal cadence SphinX spectra. $T$ and $EM$, derived in such a way vary in the range of 2.7-15.7 MK and 1.15-28.66 ($\times 10^{47}$ $cm^{-3}$), respectively. Next, we derive the thermodynamic measure ($\eta$, see \citet{Sylwester1995}, \citet{Sylwester2006}) which is associated with the thermal energy as follows:

\begin{equation}
\label{eth}
 E_{th}=3 k_b\eta \sqrt{V}
\end{equation}
Here, $V$ is the volume of the emitting plasma. Thermodynamic measure, $\eta$, defined as $T\sqrt{EM}$, characterizes the thermal energy of the plasma for the case of the constant volume of the emitting region. In this study, we derive the volume of the emitting region from the EUV images in 284 $\mathrm{\AA}$, the hottest channel, as obtained from \textit{STEREO} twin satellites. Figure \ref{st-a-b-eit-contour} shows the sequence of images in 284 $\mathrm{\AA}$ during the flare, recorded by \textit{STEREO} as well as in 195 $\mathrm{\AA}$ by EIT/\textit{SOHO}. The contours drawn on the images are 5, 10 and 20\% of the maximum intensity of respective images.

\begin{figure}[!htbp]
  \begin{center}
   \includegraphics[width=0.45\textwidth, height=0.45\textwidth, angle=0]{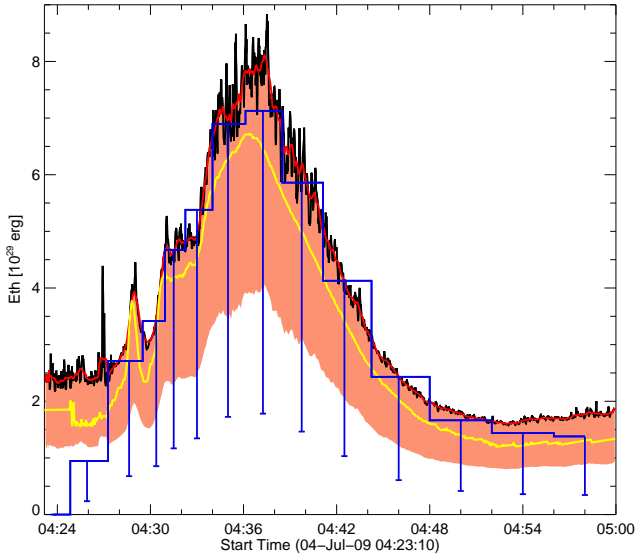}
   \caption{Temporal evolution of the iso-thermal and multi-thermal energy content during the flare. Iso-thermal energy, estimated from SphinX observations, is plotted in black (smoothed in red) while that derived from \textit{GOES} observations in shown in yellow color. Multi-thermal energy, derived by applying W-S algorithm on the SphinX \& SOXS combined data is shown in the form of blue color histogram. The uncertainty in the estimation of iso-thermal energy content is shown by filled area (light red) while the same corresponding to the multi-thermal energy content is shown in the form of error bars (blue).}
   \label{E-th}
 \end{center}
\end{figure}

From Figure \ref{st-a-b-eit-contour}, it may be noted that the contour of green color, corresponding to 20\% of the maximum intensity best represents the emitting region. Volume of the region, assuming spherical geometry, is estimated as follows:

\begin{equation}
\label{eq-volume}
V=\frac{4}{3} \pi (R)^3
\end{equation}
where `$R$' is the equivalent radius [=$(A/\pi)^{1/2}$] of a circle having area ($A$) equal to that of the region within the iso-contour of 20\% of the maximum intensity (green color) of the images presented in Figure \ref{st-a-b-eit-contour}. We estimate the temporal evolution of the flare volume from the images in several EUV wavelengths (171, 195 and 284 $\mathrm{\AA}$) made available by \textit{STEREO} and EIT/\textit{SOHO} satellites, in the aforesaid manner. However, the volume estimated from the images of 284 $\mathrm{\AA}$ (representing the hot plasma region) is used for deriving thermal energetics of the flare. As \textit{STEREO} provides the images of the region with a time cadence of 20 minutes, we interpolate the flare volume at intermediate times using cubic spline interpolation technique. The flare volume, derived in the aforesaid manner, varies in the range of 1.2-5.4 ($\times 10^{28}$ $cm^3$). Using the volume estimated above, we derive the time evolution of the iso-thermal energy content ($E_{th}$) during the flare using Equation \ref{eth} as plotted (black color) in Figure \ref{E-th}. The iso-thermal energy content estimated in such a way varies in the range of 2-9 $\times 10^{29}$ ergs. In a similar fashion, we also estimate the iso-thermal energy content employing the $T$ and $EM$ derived from the \textit{GOES} observations (cf. Figure \ref{x-ray-sphinx-soxs-goes-lc}). Thus the iso-thermal energy, as derived from the \textit{GOES} observations, is found to be varying in the range of 1.5-6.5$\times 10^{29}$ ergs during the flare and shown by yellow color plot in the Figure \ref{E-th}.

The estimation of source size may contain various kinds of uncertainties which is crucial to investigate as it is subsequently propagated to the thermal energy estimates. As the EUV source sizes are used with the DEM[T] distribution (derived from X-ray observations) while estimating the thermal energy content, a disagreement between the co-temporal source sizes within EUV and X-ray waveband may become one of the major contributor to the uncertainty. Unfortunately imaging mode observations in X-ray waveband have not been available for this flare, and hence EUV images have been used in this study for source size estimation. Although the 284 $\mathrm{\AA}$  filter provides the peak temperature response (at $\sim$2MK), maximum among the other EUV wavelengths available from \textit{STEREO} satellites, it is still quite far from flare plasma temperatures in which X-ray emission is obtained. In this regard, we estimate the co-temporal X-ray and EUV source sizes of thirteen flares of intensity class B1.1 - C1.0 which have occurred during July 04-06, 2009 in the active region AR11024. It may be noted that SOL2009-07-04T04:37, the flare of our present study, is also produced from the same active region. In this statistical investigation, we have used X-ray images obtained from \textit{HINODE}/XRT while EUV images \textit{STEREO} twin satellites in 195 $\mathrm{\AA}$. The source size in both the aforesaid wavelengths is estimated employing the same approach as discussed previously. The comparative investigation has revealed that the source size estimated from the X-ray images are systematically smaller than that derived from the EUV images whereas the ratio is varying in the range of 1.1-9.0 with a median value of 6. Next, imaging a asymmetric flaring region with the instruments that observe the Sun from different angles e.g. the observation of AR11024 with \textit{STEREO}-A and B satellites (Figure \ref{stereo-twin}), may contribute to additional uncertainty in the source size estimation. In view of the same, we have also made a comparison of EUV source sizes estimated from 195 $\mathrm{\AA}$  images obtained from \textit{STEREO}-A and B satellites for the aforesaid thirteen flares. This study revealed that the orthogonal view of the flaring region systematically results in larger source size by a factor varying in the range of 1.2-1.5 than that calculated from the images having on-disk view. As the uncertainty in the source size, arising due to the difference in EUV and X-ray source sizes is larger than than that occurred due to observing the region in different angles, the latter may be neglected while calculating the uncertainties in the thermal energy estimates. Thus, in conclusion, considering the fact that EUV source sizes are systematically larger than X-ray sources by a factor of 6, the volume derived from the same suffers an overestimation by a factor of 4. Employing the scheme of error propagation, the application of aforesaid uncertainty in the volume estimates may result in the overestimation of thermal energy content (Equation \ref{eq-volume}) by a factor of $\sim$2. We show the aforesaid uncertainty in the iso-thermal energy estimate in the form of associated filled area (light red color) in Figure \ref{E-th}.

Next, we estimate the multi-thermal energy content of the flare with the help of the DEM[T] distribution, derived from the W-S inversion scheme, as per the following equation \citep{Sylwester2014,Aschwanden2015}:

\begin{equation}
\label{eq-multi-thermal-energy}
E_{th}=3 k_B V^{1/2} \sum_k T_k DEM_k^{1/2}
\end{equation}

The multi-thermal energy content derived in the aforesaid manner varies in the range of 1-7$\times 10^{29}$ ergs as plotted in the form of blue color histogram in Figure \ref{E-th}. In the estimation of multi-thermal energetics, we have employed the combined data-set which is prepared from the X-ray emission in 1.6-5.0 keV as obtained from SphinX and in 5.0-8.0 keV (with the application of a normalization factor of `2.5') from SOXS (cf. Section \ref{sec:data-analysis}). It may be argued that this scheme of normalization is biased. In this regard, we made a parallel case study in which the best-fit DEM[T] distribution is derived using combined data-set which, however, is prepared by applying the inverse normalization factor on the SphinX observations while considering SOXS observations to be true. This investigation resulted in the DEM values systematically lowered by a factor of 2.5 in comparison to that estimated in the previous case. On the other hand, the best-fit plasma temperature values remain unchanged (also see \citet{Mrozek2012}). Therefore, considering the fact that DEM values in the former case are larger by the aforesaid factor i.e. 2.5, multi-thermal energy content is resulted to be overestimated by a factor of 4 (Equation \ref{eq-multi-thermal-energy}). In this calculation, we have also included the uncertainty in the volume estimation obtained previously. In Figure \ref{E-th}, we show the uncertainty in the multi-thermal energy estimates (blue) with the low error bars.

The comparison of iso-thermal and multi-thermal energy content for this flare (Figure \ref{E-th}) revealed that multi-thermal energy matches well with the iso-thermal energy during the rise and the decay phase of the flare. However, during the maximum of the impulsive phase, minor disagreement in the form of lower values of multi-thermal energy in comparison to the iso-thermal energy is noted.

Next, we derive the multi-thermal energy from the best-fit DEM[T] distribution, obtained by employing various DEM schemes on the observed SXR spectrum in 1.6-5.0 keV (low-energy), 5.0-8.0 keV (high-energy) and 1.6-8.0 keV (combined-energy) bands, during the peak of impulsive phase of the flare (04:36:00-04:38:30 UT). Multi-thermal energy for the low-energy band SXR is estimated to be 177, 225 and 4.1 $\times 10^{29}$ ergs, corresponding to the the single-gaussian, power-law and W-S DEM schemes, respectively. On the other hand, the energy content derived by employing the aforesaid DEM schemes on the high-energy band SXR is obtained to be 64, 65 and 1.5 $\times 10^{29}$ ergs, respectively. Further, the multi-thermal energy is resulted to be 85, 91 and 4.4 $\times 10^{29}$ ergs by applying the aforesaid DEM schemes on the combined-energy band SXR, respectively. By comparing the above mentioned energy estimates, we note that the flare energetics, estimated from the parameters derived only from the spectral inversion of the low-energy band of SXR spectrum leads to higher values than that obtained from combined-energy band case. On the other hand, the multi-thermal energies, resulted by applying various DEM schemes on the high-energy part of SXR spectrum, are estimated to be lower than that obtained from combined-energy band SXR spectrum. This trend is consistently noted in the energetics estimated employing all the aforesaid DEM schemes. On the contrary, we find that the best-fit DEM[T] distribution, obtained with the DEM schemes which postulate either the single gaussian or power-law functional dependence of DEM, leads to the overestimation of multi-thermal energy by approximately one order in comparison to that estimated from W-S algorithm.

\section{Summary and Conclusions}
\label{sec:sum-conc}
We investigate the thermal characteristics of the flare plasma by analysing X-ray emission in the energy band 1.6-8.0 keV observed during SOL2009-07-04T04:37 flare, the only common event observed by SphinX and SOXS instruments. We derive the evolution of the best-fit DEM[T] distribution during the flare by employing various DEM inversion algorithms. In addition, we have also studied the dependence of the best-fit DEM[T] corresponding to various input energy bands within SXR emission. Following are the key points of our study:
\begin{enumerate}
\item Best-fit DEM[T] distribution for the low (1.6-5.0 keV), high (5.0-8.0 keV) and combined-energy band (1.6-8.0 keV) of X-ray emission during the flare resulted in  higher values of $DEM_p$, however at low $T_p$ for the low-energy band in comparison to the relatively lower values of $DEM_p$ at higher $T_p$ obtained by analyzing the high-energy band of the SXR.
\item We derive the time evolution of DEM[T] distribution during various phases of the flare by employing Withbroe-Sylwester maximum likelihood DEM inversion algorithm on the individual as well as combined observations of SphinX and SOXS during the flare. The results are summarised as follows:
\begin{itemize}
  \item The best-fit DEM[T] distribution corresponding to the X-ray emission during the flare onset can be well represented by a single gaussian function with a width of $\sim$1 MK, which suggests flare plasma to be of iso-thermal nature in this phase.
  \item Analysis of X-ray emission during the rise to the peak of impulsive phase of the flare revealed the presence of multi-thermal plasma as the corresponding best-fit DEM[T] curves show double gaussian form with the widths of $\sim$ 1.5 MK.
  \item Temporal evolution of the best-fit DEM[T] distribution corresponding to the post-maximum phase of the flare can be well represented by a single gaussian function, however, with the peak temperature varying in the range of $\sim$ 13.0-5.5 MK.
\end{itemize}
\item Iso-thermal and multi-thermal energy content is estimated during the flare. We find that the multi-thermal energy estimates are in close agreement with the iso-thermal energy values except during the peak of the impulsive phase of the flare where iso-thermal energy is estimated to be larger than the multi-thermal energy content.
\item Multi-thermal energy is determined from the best-fit DEM[T] distribution resulted from the application of various inversion schemes on the X-ray emission measured during the peak of the impulsive phase of the flare. We find that the energy content estimated from the parameters derived only from spectral inversion of the low-energy band (1.6-5.0 keV) of SXR spectrum result in larger values than that obtained from the analysis of the SXR emission in combined-energy band. On the contrary, the same derived from only the high-energy band of SXR spectrum leads to lower estimates when comparing with the energy values calculated from combined-energy band analysis. This trend is consistently resulted in the thermal energetics determined from all the DEM schemes. This suggests that the observations of SXR emission during a flare in the combined-energy band with high temporal and energy cadence is very important to derive the complete thermal energetics of the flare.
\item The best-fit DEM[T] distribution obtained for the DEM schemes which postulate either single-gaussian or power-law functional form of DEM-T curve, lead to the estimation of thermal energy content much higher by approximately one order than that estimated from the W-S scheme. This can be understood by the fact that the width of the best-fit DEM[T] distribution, obtained by employing single-gaussian approach (see Figure \ref{fit-dem-gauss-spectra}) is larger than that resulted by the application of W-S scheme (Figure \ref{w-s-dem-fit-sphinx-soxs}). It may be noted that this disagreement between various DEM inversion schemes, and hence thermal energy estimates can have significant impact in the context of coronal heating from low intensity class (micro- and nano-) flares. However, as X-ray emission covers only high-temperature corona, recent studies focussing coronal heating energized by small intensity flares also combine multi-wavelength observations with the X-ray emission during flares \citep{Testa2014}. Moreover, several advanced schemes of DEM inversion viz. `DEM\_manual' \citep{Schmelz2015}, `EM Loci approach' \citep{Cirtain2007}, combination of gaussian and power-law functional form of DEM \citep{Guennou2013, Aschwanden2015SoPh} etc. have also been employed in deriving thermal characteristics of emission measure during small intensity class flares. Therefore, in future, we plan to extrapolate the application of W-S DEM inversion scheme on the combined EUV and X-ray observations during small flares in order to make a comparative survey of thermal energy content derived by W-S method and other DEM inversion schemes.
\end{enumerate}

\acknowledgements
This research has been supported by Polish NCN grant 2011/01/B/ST9/05861 and from the European Commission's Seventh Framework Programme under the grant agreement No. 284461 (eHEROES project). Moreover, the research leading to these results has received funding from the European Community's Seventh Framework Programme (FP7/2007-2013) under grant agreement no. 606862 (F-CHROMA). Authors also acknowledge the open data policy of the SphinX, SOXS, \textit{SOHO}, \textit{HINODE} and \textit{STEREO} missions. SAO/ADS abstract service is duly acknowledged for providing the up-to-date and well-organized bibliography. Additionally, the Coyote's IDL programming support is acknowledged. Authors also thank the anonymous referee for constructive comments which improved the manuscript.

\clearpage


\begin{thebibliography}{}

\bibitem[Aschwanden(2007)]{Aschwanden2007} Aschwanden, M.~J.\ 2007,
\apj, 661, 1242

\bibitem[Aschwanden et al.(2015)]{Aschwanden2015} Aschwanden, M.~J.,
Boerner, P., Ryan, D., et al.\ 2015, \apj, 802, 53

\bibitem[Aschwanden et al.(2015)]{Aschwanden2015SoPh} Aschwanden, M.~J., Boerner, P., Caspi, A., et al.\ 2015, \solphys, 290, 2733

\bibitem[Aschwanden et al.(2014)]{Aschwanden2014} Aschwanden, M.~J.,
Xu, Y., \& Jing, J.\ 2014, \apj, 797, 50

\bibitem[Awasthi et al.(2014)]{Awasthi2014} Awasthi, A.~K., Jain,
R., Gadhiya, P.~D., et al.\ 2014, \mnras, 437, 2249

\bibitem[Benz(2008)]{Benz2008} Benz, A.~O.\ 2008, Living Reviews
in Solar Physics, 5, 1

\bibitem[Brown(1971)]{Brown1971} Brown, J.~C.\ 1971, \solphys,
18, 489

\bibitem[Caspi
\& Lin(2010)]{Caspi2010} Caspi, A., \& Lin, R.~P.\ 2010, \apjl, 725, L161

\bibitem[Choudhary et al.(2013)]{Choudhary2013} Choudhary, D.~P.,
Gosain, S., Gopalswamy, N., et al.\ 2013, Advances in Space Research, 52,
1561

\bibitem[Cirtain et al.(2007)]{Cirtain2007} Cirtain, J.~W., Del Zanna, G., DeLuca, E.~E., et al.\ 2007, \apj, 655, 598

\bibitem[Craig \& Brown(1976)]{Craig1976} Craig, I.~J.~D., \& Brown, J.~C.\ 1976, \aap, 49, 239

\bibitem[Dalmasse et al.(2015)]{Dalmasse2015} Dalmasse, K., Chandra, R., Schmieder, B., \& Aulanier, G.\ 2015, \aap, 574, A37

\bibitem[Delaboudini{\`e}re et al.(1995)]{Delaboudiniere1995}
Delaboudini{\`e}re, J.-P., Artzner, G.~E., Brunaud, J., et al.\ 1995,
\solphys, 162, 291

\bibitem[Del Zanna et al.(2015)]{Del Zanna2015} Del Zanna, G., Dere, K.~P., Young, P.~R., Landi, E., \& Mason, H.~E.\ 2015, \aap, 582, A56

\bibitem[Dere \& Cook(1979)]{Dere1979} Dere, K.~P., \& Cook, J.~W.\ 1979, \apj, 229, 772

\bibitem[Fletcher et al.(2011)]{Fletcher2011} Fletcher, L., Dennis,
B.~R., Hudson, H.~S., et al.\ 2011, \ssr, 159, 19

\bibitem[Gburek et al.(2011)]{Gburek2011} Gburek, S., Sylwester,
J., Kowalinski, M., et al.\ 2011, Solar System Research, 45, 189

\bibitem[Gburek et al.(2013)]{Gburek2013} Gburek, S., Sylwester,
J., Kowalinski, M., et al.\ 2013, \solphys, 283, 631

\bibitem[Guennou et al.(2013)]{Guennou2013} Guennou, C., Auch{\`e}re, F., Klimchuk, J.~A., Bocchialini, K., \& Parenti, S.\ 2013, \apj, 774, 31

\bibitem[Jain et al.(2008)]{Jain2008} Jain, R., Aggarwal, M.,
\& Sharma, R.\ 2008, Journal of Astrophysics and Astronomy, 29, 125

\bibitem[Jain et al.(2011a)]{Jain2011a} Jain, R., Awasthi, A.~K.,
Chandel, B., et al.\ 2011a, \solphys, 271, 57

\bibitem[Jain et al.(2011b)]{Jain2011b} Jain, R., Awasthi, A.~K.,
Rajpurohit, A.~S., \& Aschwanden, M.~J.\ 2011b, \solphys, 270, 137

\bibitem[Jain et al.(2005)]{Jain2005} Jain, R., Dave, H., Shah,
A.~B., et al.\ 2005, \solphys, 227, 89


\bibitem[Kepa et al.(2008)]{Kepa2008} Kepa, A., Sylwester, B.,
Siarkowski, M., \& Sylwester, J.\ 2008, Advances in Space Research, 42, 828

\bibitem[Kepa et al.(2006)]{Kepa2006} Kepa, A., Sylwester, J.,
Sylwester, B., Siarkowski, M.,
\& Stepanov, A.~I.\ 2006, Solar System Research, 40, 294

\bibitem[Kulinov{\'a} et
al.(2011)]{Kulinova2011} Kulinov{\'a}, A., Ka{\v s}parov{\'a}, J., Dzif{\v c}{\'a}kov{\'a}, E., et al.\ 2011, \aap, 533, A81

\bibitem[Landi et al.(2012)]{Landi2012} Landi, E., Del Zanna, G.,
Young, P.~R., Dere, K.~P., \& Mason, H.~E.\ 2012, \apj, 744, 99

\bibitem[Li et
al.(2005)]{Li2005} Li, H., Berlicki, A., \& Schmieder, B.\ 2005, \aap, 438, 325

\bibitem[Mrozek et al.(2012)]{Mrozek2012} Mrozek, T., Gburek, S.,
Siarkowski, M., et al.\ 2012, Central European Astrophysical Bulletin, 36,
71

\bibitem[Saint-Hilaire
\& Benz(2005)]{Saint-Hilaire2005} Saint-Hilaire, P., \& Benz, A.~O.\ 2005, \aap, 435, 743

\bibitem[Schmelz \& Winebarger(2015)]{Schmelz2015} Schmelz, J.~T., \& Winebarger, A.~R.\ 2015, Philosophical Transactions of the Royal Society of London Series A, 373, 20140257

\bibitem[Shibata(1999)]{Shibata1999} Shibata, K.\ 1999, \apss, 264, 129


\bibitem[Sylwester et al.(2006)]{Sylwester2006} Sylwester, B.,
Sylwester, J., Kepa, A., et al.\ 2006, Solar System Research, 40, 125

\bibitem[Sylwester et al.(2014)]{Sylwester2014} Sylwester, B.,
Sylwester, J., Phillips, K.~J.~H., K{\c e}pa, A.,
\& Mrozek, T.\ 2014, \apj, 787, 122


\bibitem[Sylwester et
al.(1995)]{Sylwester1995} Sylwester, J., Garcia, H.~A., \& Sylwester, B.\ 1995, \aap, 293, 577

\bibitem[Sylwester et al.(2012)]{Sylwester2012} Sylwester, J.,
Kowalinski, M., Gburek, S., et al.\ 2012, \apj, 751, 111

\bibitem[Sylwester et al.(1980)]{Sylwester1980} Sylwester, J.,
Schrijver, J., \& Mewe, R.\ 1980, \solphys, 67, 285

\bibitem[Testa et al.(2014)]{Testa2014} Testa, P., De Pontieu, B., Allred, J., et al.\ 2014, Science, 346, 1255724


\end{thebibliography}
\end{document}